\documentclass[sigplan,screen,9pt]{acmart}

\usepackage{algorithmic}
\usepackage{graphicx}
\usepackage{textcomp}
\usepackage{multirow}
\usepackage{makecell}
\usepackage{booktabs}
\usepackage{listings}
\usepackage{tikz}
\usepackage{pifont}
\usepackage{comment}
\usepackage{enumitem}
\usepackage{scalerel}
\usepackage[pages=some]{background}

\makeatletter
\let\OriginalBaselineStretch\ACM@origbaselinestretch
\makeatother

\definecolor{Lavender}{HTML}{e6e6fa}
\definecolor{WhiteSmoke}{HTML}{f5f5f5}

\usepackage[all=normal,paragraphs=tight,floats=tight,mathspacing=tight,wordspacing=tight,leading=tight,mathdisplays=tight]{savetrees}
\usepackage{setspace}
\microtypesetup{stretch=0,shrink=125}

\definecolor{codegreen}{rgb}{0,0.6,0}
\definecolor{codegray}{rgb}{0.5,0.5,0.5}
\definecolor{codepurple}{rgb}{0.58,0,0.82}
\definecolor{codebgcolor}{rgb}{0.95,0.95,0.92}
\lstdefinestyle{codestyle}{
    backgroundcolor=\color{codebgcolor},   
    commentstyle=\color{codegreen},
    keywordstyle=\color{magenta},
    numberstyle=\tiny\color{codegray},
    stringstyle=\color{codepurple},
    basicstyle=\ttfamily\footnotesize,
    breakatwhitespace=false,         
    breaklines=true,                 
    captionpos=b,                    
    keepspaces=true,                 
    showspaces=false,                
    showstringspaces=false,
    showtabs=false,                  
    tabsize=2,
    escapechar=\%,
    columns=fullflexible,
}
\newcommand*\emptycirc[1][1ex]{\tikz\draw (0,0) circle (#1);} 
\newcommand*\halfcirc[1][1ex]{%
  \begin{tikzpicture}
  \draw[fill] (0,0)-- (90:#1) arc (90:270:#1) -- cycle ;
  \draw (0,0) circle (#1);
  \end{tikzpicture}}
\newcommand*\fullcirc[1][1ex]{\tikz\fill (0,0) circle (#1);}

\newcommand{\highlightedtt}[1]{%
  \smash{\setlength{\fboxsep}{0.5pt}\colorbox{WhiteSmoke!70!Lavender}{\texttt{#1}}}%
}

\newcommand{\highlightedurl}[1]{%
  \href{#1}{\texttt{\textcolor{blue}{#1}}}%
}

\newcommand{\highlightedurltext}[2]{%
  \href{#1}{\texttt{\textcolor{blue}{#2}}}%
}

\copyrightyear{2024}
\acmYear{2024}
\setcopyright{rightsretained}
\acmConference[MLCAD '24]{2024 ACM/IEEE International Symposium on
Machine Learning for CAD}{September 9--11, 2024}{Salt Lake City, UT, USA}
\acmBooktitle{2024 ACM/IEEE International Symposium on Machine Learning for
CAD (MLCAD '24), September 9--11, 2024, Salt Lake City, UT, USA}
\acmDOI{10.1145/3670474.3685961}
\acmISBN{979-8-4007-0699-8/24/09}

\newcommand{\ourwork}{{HLSFactory}}

\begin{document}

\title{\ourwork: A Framework Empowering High-Level Synthesis Datasets for Machine Learning and Beyond}
\author{
    Stefan Abi-Karam\textsuperscript{1,2},
    Rishov Sarkar\textsuperscript{1},
    Allison Seigler\textsuperscript{3},
    Sean Lowe\textsuperscript{4},
    Zhigang Wei\textsuperscript{3},
    Hanqiu Chen\textsuperscript{1},\\
    Nanditha Rao\textsuperscript{5},
    Lizy John\textsuperscript{3},
    Aman Arora\textsuperscript{4},
    Cong Hao\textsuperscript{1}
}
\affiliation{
    \textsuperscript{1}Georgia Institute of Technology, 
    \textsuperscript{2}Georgia Tech Research Institute, 
    \textsuperscript{3}The University of Texas at Austin,\\
    \textsuperscript{4}Arizona State University, 
    \textsuperscript{5}International Institute of Information Technology Bangalore\\
    { \normalsize
    \{%
        \href{mailto:stefanabikaram@gatech.edu}{\nolinkurl{stefanabikaram}},
        \href{mailto:rishov.sarkar@gatech.edu}{\nolinkurl{rishov.sarkar}},
        \href{mailto:hanqiu.chen@gatech.edu}{\nolinkurl{hanqiu.chen}},
        \href{mailto:callie.hao@gatech.edu}{\nolinkurl{callie.hao}}\}@gatech.edu,
    \{%
        \href{mailto:aseigler@utexas.edu}{\nolinkurl{aseigler}},
        \href{mailto:zw5259@utexas.edu}{\nolinkurl{zw5259}},
        \href{mailto:ljohn@ece.utexas.edu}{\nolinkurl{ljohn@ece.utexas.edu}}\}@utexas.edu,
    \\
    \{%
        \href{mailto:slowe8@asu.edu}{\nolinkurl{slowe8}},
        \href{mailto:aman.kbm@asu.edu}{\nolinkurl{aman.kbm}}\}@asu.edu,
    \{%
        \href{mailto:nanditha.rao@iiitb.ac.in}{\nolinkurl{nanditha.rao}}\}@iiitb.ac.in
    }
    \country{}
}
\date{}

\begin{abstract}

Machine learning (ML) techniques have been applied to high-level synthesis (HLS) flows for quality-of-result (QoR) prediction and design space exploration (DSE). Nevertheless, the scarcity of accessible high-quality HLS datasets and the complexity of building such datasets present great challenges to FPGA and ML researchers. Existing datasets either cover only a subset of previously published benchmarks, provide no way to enumerate optimization design spaces, are limited to a specific vendor, or have no reproducible and extensible software for dataset construction. Many works also lack user-friendly ways to add more designs to existing datasets, limiting wider adoption and sustainability of such datasets. 

In response to these challenges, we introduce HLSFactory, a comprehensive framework designed to facilitate the curation and generation of high-quality HLS design datasets. HLSFactory has three main stages: 1) a design space expansion stage to elaborate single HLS designs into large design spaces using various optimization directives across multiple vendor tools, 2) a design synthesis stage to execute HLS and FPGA tool flows concurrently across designs, and 3) a data aggregation stage for extracting standardized data into packaged datasets for ML usage. This tripartite architecture not only ensures broad coverage of data points via design space expansion but also supports interoperability with tools from multiple vendors.  Users can contribute to each stage easily by submitting their own HLS designs or synthesis results via provided user APIs. The framework is also flexible, allowing extensions at every step via user APIs with custom frontends, synthesis tools, and scripts.

To demonstrate the framework functionality, we include an initial set of built-in base designs from PolyBench, MachSuite, Rosetta, CHStone, Kastner et al.’s Parallel Programming for FPGAs, and curated kernels from existing open-source HLS designs. We report the statistical analyses and design space visualizations to demonstrate the completed end-to-end compilation flow, and to highlight the effectiveness of our design space expansion beyond the initial base dataset, which greatly contributes to dataset diversity and coverage.

In addition to its evident application in ML, we showcase the versatility and multi-functionality of our framework through seven case studies: I) Building an ML model for post-implementation QoR prediction; II) Using design space sampling in stage 1 to expand the design space covered from a small base set of HLS designs; III) Demonstrating the speedup from the fine-grained design parallelism backend; IV) Extending HLSFactory to target Intel's HLS flow across all stages; V) Adding and running new auxiliary designs using HLSFactory; VI) Integration of previously published HLS data in stage 3; VII) Using HLSFactory to perform HLS tool version regression benchmarking.

Code available at \textcolor{blue}{\url{https://github.com/sharc-lab/HLSFactory}}.
\end{abstract}

\maketitle
\backgroundsetup{opacity=1, scale=1, angle=0, contents={
\begin{tikzpicture}[remember picture, overlay]
\node[anchor=north east, inner xsep=50pt, inner ysep=10pt] at (current page.north east) {
\href{https://www.acm.org/publications/policies/artifact-review-and-badging-current}{
\includegraphics[width=40pt]{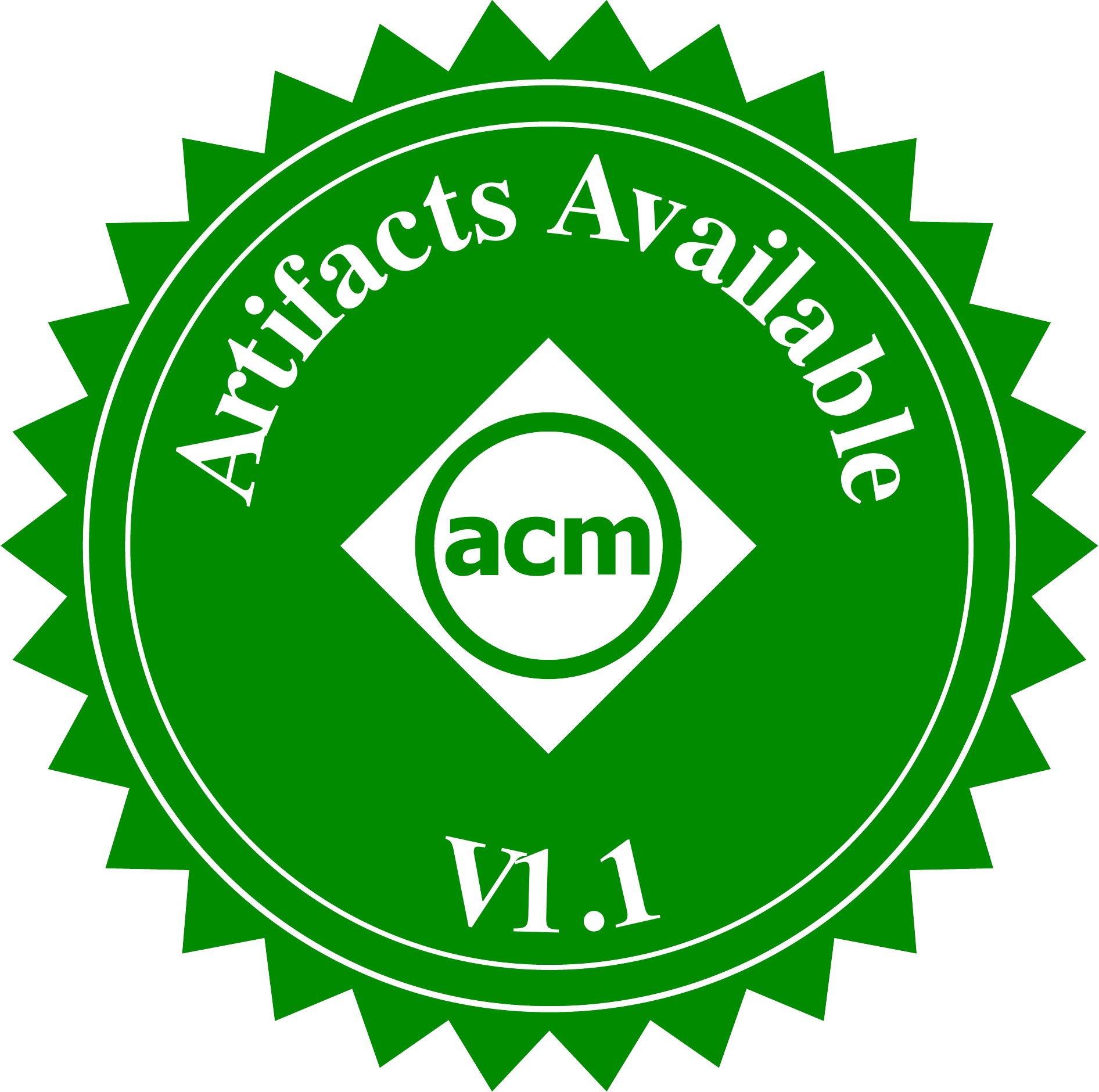}
\includegraphics[width=40pt]{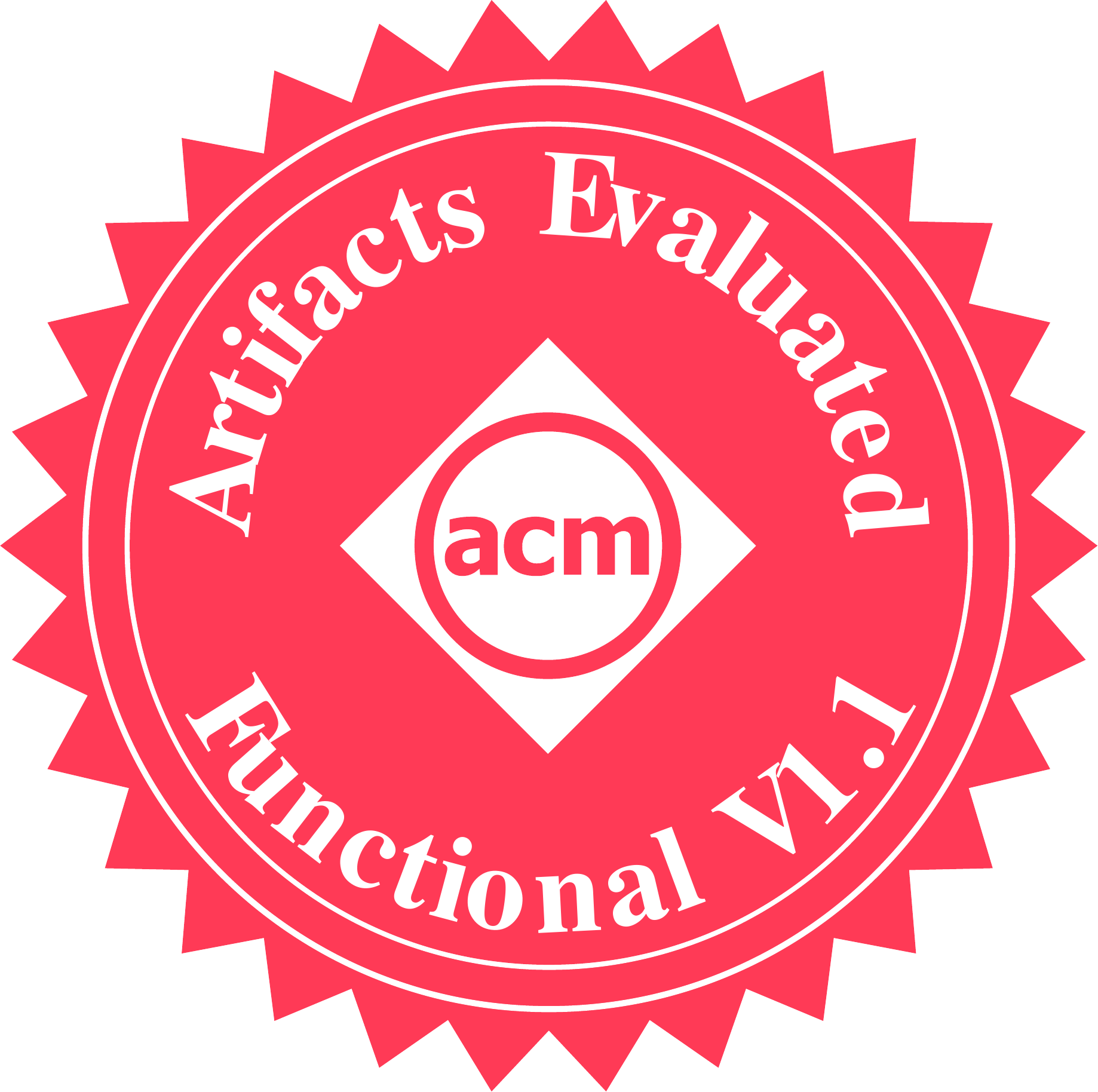}
}};
\end{tikzpicture}
}}
\BgThispage

\pagestyle{plain}

\section{Introduction}

Machine learning (ML) techniques have been widely applied to different  electronic design automation (EDA) flows including high-level synthesis (HLS) ~\cite{hls_pyramid,dai2018fast,ML_DSE,ML_iscas1,ml_cgra,ml_eda1,ml_eda2, HL-pow, LEAPER, PowerGear} for quality-of-result (QoR) prediction, optimization, and design space exploration (DSE).
A key enabler to the success of such ML techniques is high-quality datasets, most of which are developed for individual studies e.g., \cite{hls_pyramid, LEAPER, HL-pow}. Some recent works have contributed open-source datasets that can be used by other reseachers, e.g., ~\cite{gautier2016spector, rosetta, hlsdataset, db4hls,mlsbench, chang2023dr}.

Despite the great benefits of these datasets, 
there are still fundamental \textbf{limitations} that hinder their wider adoption for ML applications and FPGA research.
\underline{First}, these datasets are usually small or homogeneous, containing only a subset of previously published HLS benchmarks \cite{polybench, chstone,machsuite,rosetta}, and frequently consisting exclusively of designs that work with one HLS tool from a single vendor.
For example, Spector~\cite{gautier2016spector} contains only 9 Intel HLS designs, HLSDataset~\cite{hlsdataset} contains 34 AMD/Xilinx HLS designs, and Rosetta~\cite{rosetta} contains 6 AMD/Xilinx HLS designs.
\underline{Second}, because of these separately developed HLS datasets, the designs and intermediate/final tool outputs, which serve as important ML model features, are often reported organized in non-standard \textit{ad hoc} ways. Some datasets contain only source code~\cite{chstone,polybench,machsuite}, some datasets contain only resource usage and end-to-end throughput~\cite{rosetta} but no clock frequency or power numbers,
while some contain only post-implementation results~\cite{chang2023dr}. HLSyn~\cite{chang2023dr} is a dataset for HLS designs targeted towards predicting design quality of FPGAs. It consists of a wider range of programs and compiler directives, enabling performance optimization of designs. 
However, existing datasets require huge manual effort and deep domain-specific knowledge for ML practitioners if they need a complete, unified, and larger dataset, where they must execute all related HLS tools on their own to re-collect and organize the needed information.
\underline{Third}, it is challenging for external users who want to extend the existing datasets by contributing their own designs, primarily caused by ad-hoc data formats and missing details when building these datasets (e.g., tool version, target FPGA device, clock frequency, implementation flow settings).
\textbf{The fundamental limitation is, however, not the lack of another complete and rich HLS dataset, but rather the lack of a flexible and extensible framework to enable continuous contributions to a standardized and sustainable dataset.}

Therefore, in this work, we introduce \textbf{\ourwork}, the first framework that takes a principled approach to HLS dataset generation, collection, expansion, and integration, aiming to facilitate a continuous and community-wide effort to contribute to the richest HLS dataset, which will keep expanding easily. \ourwork{} boasts the following features:

\setlist[itemize]{leftmargin=1em}
\begin{itemize}
    \item \textbf{Complete and easily extensible with user inputs at multiple stages.} 
    \ourwork{} has an end-to-end compilation flow including three main stages: design space expansion stage to elaborate single HLS designs at the source-code level into large design spaces; design synthesis stage to execute HLS and FPGA tools; and data aggregation stage for extracting standardized data organization. 
    \ourwork{} uses a modular design that allows users to plug in their own designs and tool flows to the dataset with minimal effort at arbitrary stages.

    \item \textbf{Diverse and comprehensive.} The initially included dataset covers a wide variety of HLS designs, containing both simple designs synthesized with AMD/Xilinx and Intel tool flows and complex designs using Xilinx-specific features.
    In addition, \ourwork{} has a novel design space expansion and sampling approach, allowing the generation of many design points from a single HLS design, improving overall design space coverage.
    Further, \ourwork{} has comprehensive data metrics from synthesis to implementation, e.g., HLS synthesis reported resource and latency, and post-implementation resource, timing, power, etc.
    
    \item \textbf{Reproducible and user-friendly.} \ourwork{} features push-button ease-of-use to run the entire end-to-end dataset generation workflow, allowing anyone to replicate our generated results, and to easily contribute to the framework and the dataset.
    Specifically, our framework makes it extremely easy for  researchers in the FPGA community to contribute data for various FPGA devices.
    
    \item \textbf{ML-ready and multi-purpose}
    Beyond simply being used for ML training, as demonstrated with a post-implementation QoR prediction ML model (case study in \S\ref{sec:qor}), \ourwork{} is useful for any task where a large, diverse set of HLS runs is needed, like HLS tool version regression testing (case study in \S\ref{sec:reg}).
    
    \item \textbf{High performance and open-source.} \ourwork{} maximizes parallelism for fast dataset generation of large numbers of designs. \ourwork{} is open-source and available on GitHub, including both the end-to-end framework and a large set of sample designs.
\end{itemize}

In Sec.~\ref{sec:related-work}, we first provide background on the prior works in existing HLS benchmarks and datasets. Sec.~\ref{sec:framework} introduces our \ourwork{} framework detailing the three stages. Sec.~\ref{sec:implementation-and-usage} dives into the implementation of \ourwork, including how it is configured and extended, and our fine-grained parallelism technique to speed up dataset generation. We then perform several case studies in Sec.~\ref{sec:results} that demonstrate the multi-purpose of the proposed framework.

\begin{table}[t]
    \scriptsize
    \centering
    \caption{A comparison of \ourwork{} with the existing work. \fullcirc{}: feature supported; \emptycirc{}: feature unsupported; \halfcirc{}: feature partially supported.}
    \vspace{-1.0em}
    \setlength\tabcolsep{3pt}
    \renewcommand{\arraystretch}{0.5}
    \begin{tabular}{l | c | c |c |c}
        \toprule

        \textbf{Contributions}         & \textbf{DB4HLS} & \textbf{HLSyn} & \textbf{HLSDataset} & \textbf{\underline{\ourwork}} \\
        \midrule
        Benchmark --- Polybench             & \emptycirc                            & \fullcirc                            &  \fullcirc  & \fullcirc               \\
        Benchmark --- MachSuite             & \fullcirc                            & \fullcirc                           &  \fullcirc   & \fullcirc               \\
        Benchmark --- Rosetta               & \emptycirc                            & \emptycirc                         &    \fullcirc    & \fullcirc               \\
        Benchmark --- CHStone               & \emptycirc                            & \emptycirc                         &    \fullcirc    & \fullcirc               \\
        Collection --- PP4FPGA              & \emptycirc                            & \emptycirc                          &   \emptycirc     & \fullcirc               \\
        Collection --- Accelerators (\S\ref{sec:case-study-2}) & \emptycirc                    &\emptycirc                              &  \emptycirc  & \fullcirc               \\
        \hline
        Post-HLS Latency & \fullcirc & \fullcirc & \emptycirc & \fullcirc               \\
        Post-HLS Resources & \fullcirc & \fullcirc &  \fullcirc & \fullcirc               \\
        Post-HLS Artifacts & \emptycirc & \emptycirc & \emptycirc  & \fullcirc               \\
        Post-Impl. Data & \emptycirc & \emptycirc & \fullcirc & \fullcirc               \\
        \hline
        HLS Optimization DSL             & \fullcirc & \emptycirc                  &         \fullcirc     & \fullcirc \\
        Fine-Grained Parallel Builds   & \halfcirc & \emptycirc                   &        \emptycirc     & \fullcirc \\
        \hline
        Xilinx HLS Support             & \fullcirc & \fullcirc                       &     \fullcirc   & \fullcirc \\
        Intel HLS Support              & \emptycirc & \emptycirc                    &      \emptycirc      & \fullcirc \\
        \hline
        User Extendable to Other Tools & \emptycirc & \emptycirc                    &     \emptycirc       & \fullcirc \\
        Programmable API               & \emptycirc & \emptycirc                  &      \emptycirc        & \fullcirc \\
        Open Source                    & \fullcirc & \fullcirc                     &     \fullcirc     & \fullcirc \\
        \bottomrule
    \end{tabular}

    \label{tab:compare-with-existing}
\end{table}

\section{Related Work}\label{sec:related-work}

HLS community has multiple standard benchmarks for assessing HLS tools including PolyBench~\cite{polybench}, CHStone~\cite{chstone}, and MachSuite~\cite{machsuite},
which in total provide around 67 benchmark designs and are far from sufficient for ML training.
Rosetta~\cite{rosetta}, Dai~\cite{dai2018fast}, MLSBench~\cite{mlsbench}, DB4HLS~\cite{db4hls}, HLSDataset~\cite{hlsdataset}, and Spector~\cite{gautier2016spector} are all recently proposed HLS datasets, where the former four use AMD/Xilinx tools and the last uses Intel tools.
MLSBench provides a sampling from different combinations of directives (pragmas) on top of CHStone and MachSuite. DB4HLS provides exhaustive design exploration on 39 designs from MachSuite with a domain-specific language (DSL) for DSE and parallelized synthesis runs.
HLSDataset aims to cover all four commonly used benchmarks (PolyBench, CHStone, MachSuite, Rosetta) with a DSL for specifying the design space to sample from. They also illustrate two ML-based case studies for post-implementation resource and power prediction.
HLSyn~\cite{chang2023dr} uses control data flow graphs (CDFGs) of compiled HLS kernels for QoR prediction using graph neural network approaches; their designs are sampled from PolyBench and MachSuite. The features of selected prior works and \ourwork{} are shown in Table~\ref{tab:compare-with-existing}.

While existing HLS datasets serve as a solid foundation for empowering ML in HLS, they are inherently limited. First, each dataset covers only a subset of commonly used HLS benchmarks, employing ad-hoc data organization, synthesis tools, configurations, and reported metrics, lacking standardization. This fragmentation makes it exceedingly difficult for ML practitioners to effectively utilize all available datasets for training without significant efforts in data reorganization and tool re-execution. Consequently, the quality of ML models is compromised, impeding the advancement of ML in HLS. Second, the lack of standardized data organization and metric reporting poses challenges to dataset extensibility and long-term sustainability, hindering broader user contributions to HLS datasets.

Therefore, rather than introducing yet another HLS dataset, the ML for HLS community urgently requires a standard, extensible, and user-friendly \textbf{framework}. Such a framework would streamline the collection, generation, elaboration, synthesis, and organization of HLS designs and data from diverse sources and community users. This would facilitate the long-term maintenance and expansion of HLS datasets. The pressing need for such a solution is the driving force behind our proposed \ourwork{}.

\section{\ourwork{} Framework}\label{sec:framework}

\begin{figure}[t]
    \centering
    \vstretch{.98}{\includegraphics[width=0.4\textwidth]{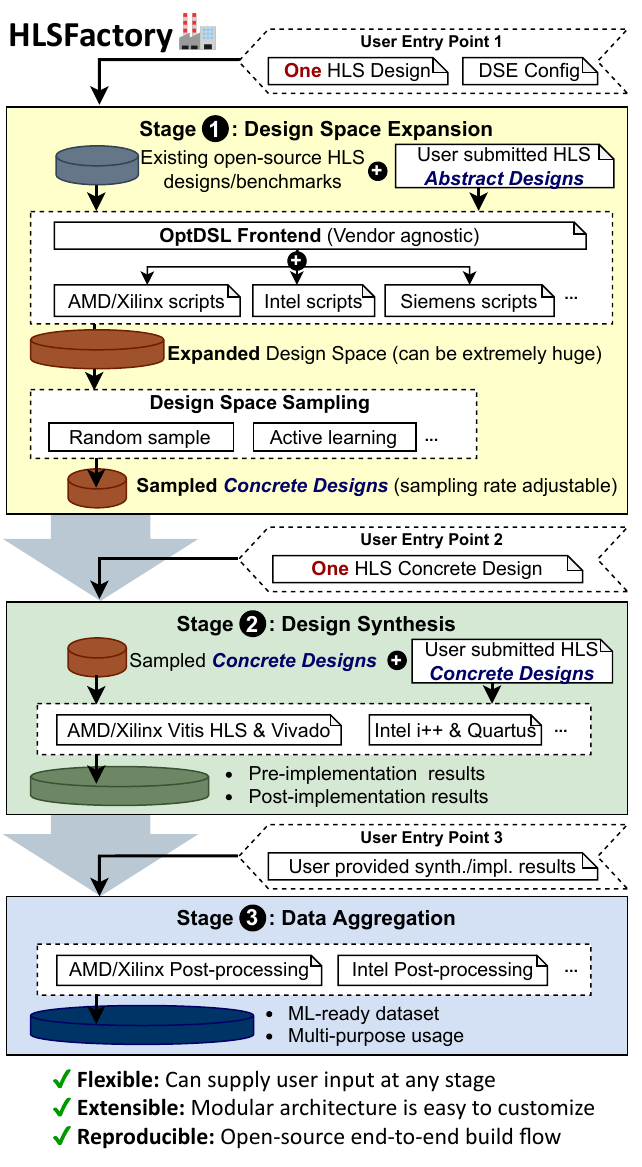}}
    \vspace{-1.5em}
    \caption{A complete overview of the \ourwork{} framework with three stages and three entry points where users can contribute their own designs.}
    \label{fig:overview}
\end{figure}

\subsection{\ourwork{} Overview}

As depicted in Fig.~\ref{fig:overview}, \ourwork{} is composed of three \textbf{stages} in its end-to-end synthesis and data extraction flow; before each stage, there is an \textbf{entry point} where users can submit designs and data.

Stage \ding{202} is \textbf{design space expansion}, aiming at expanding a single HLS design into multiple designs by enumerating different combinations of optimization directives (pragmas), which can significantly increase the number of data points for ML applications.
In this stage, users can submit one or more HLS designs with possible design space configurations and \ourwork{} will extrapolate and expand the complete design space via a frontend.
Note that the presented design space \textit{expansion} stage is explicitly different than traditional design space \textit{exploration} (commonly abbreviated as DSE). Design space expansion is not optimization guided (as detailed in \S\ref{sec:dss}), so suboptimal designs are included, broadening design space coverage needed to building robust, accurate ML models.
This frontend features multi-vendor support and allows for random sampling of generated designs to reduce the number of designs to be synthesized by HLS and implementation tools (e.g., Vivado), if needed for large design spaces, to shorten the execution time.

We will showcase this usage in Section~\ref{sec:case-study-1}.

Stage \ding{203} is \textbf{design synthesis} stage, where vendor-specific HLS and implementation tools are invoked to synthesize HLS designs into RTL code and then placed and routed. 
In this stage, users can submit their HLS designs to be directly synthesized without extrapolating. We will showcase this usage in Section~\ref{sec:case-study-2}.

Stage \ding{204} is \textbf{data aggregation}, where statistics and artifacts are collected from the implemented designs and compiled into a tool-agnostic format for use by downstream tasks such as ML training and benchmarking.
In this stage, users can submit their synthesized post-implementation results or datasets to be merged. We will showcase this usage in Section~\ref{sec:case-study-3}.

\subsection{Stage 1: Design Space Expansion and Sampling}

This stage aims at expanding a single HLS design into multiple by enumerating combinations of optimization directives (either inline or in a separate file), such as loop unroll factors, array partitioning schemes, and whether to pipeline loops. 
Such expansion is critical for ML usage because of two reasons. First, the original HLS designs and benchmarks are far from sufficient for ML training, and obtaining additional HLS designs is challenging.
Second, a key application of ML for HLS is to help designers choose the best optimization directives for their HLS designs by predicting post-HLS-synthesis and post-implementation metrics from HLS source code and directives (e.g., \cite{hlsdataset, chang2023dr,wu2021ironman,wu2022ironman}). Therefore, an ML-ready HLS dataset must provide wide coverage of how different choices of optimization directives can impact a design.
Note that the design space expansion is expected to be across \textit{multiple vendors, tools, and devices}.

On the other hand, the expanded design space can be huge, and synthesizing and implementing each design may be prohibitively time-consuming. Therefore, design space sampling is needed.

We define the concept of a \textit{frontend} pass, which \textit{lowers} an HLS \textit{abstract design} to a certain number of \textit{concrete designs}.
An HLS abstract design is not directly synthesizable but contains parameterized directives that require preprocessing.
An HLS concrete design is a copy of the abstract HLS design and is augmented with one possible combination of optimization directives from the design space.

\subsubsection{Vendor-agnostic OptDSL Frontend}\label{sec:opt-dsl-frontend}

\begin{figure}[t]
    \centering
    \begin{lstlisting}[
    basicstyle=\scriptsize\selectfont\ttfamily, 
    language=Tcl
    ]
loop_opt,3,2
0,lp2,pipeline,unroll,[1 2 4 8]
1,lp3,pipeline,unroll,[1 2 4 8]
2,lp3,,unroll,[1 2 4 8]
set_directive_unroll -factor [factor] k2mm/[name]
set_directive_pipeline k2mm/[name]
\end{lstlisting}
    \vspace{-1.0em}
    \caption{A snippet demonstrating the OptDSL syntax.}
    \label{listing:opt-dsl}
\end{figure}

For design space expansion, all possible combinations of optimization directives for a certain HLS design must be explicitly specified.
We propose a frontend using a domain-specific language (DSL), named OptDSL, to specify the design space using a DSE configuration file.
Fig.~\ref{listing:opt-dsl} shows an example of the OptDSL syntax, which specifies the choices for how to pipeline or unroll two loops \texttt{lp2} and \texttt{lp3}.

OptDSL is vendor-agnostic but based on a modified version of a Vitis HLS Tcl script, minimizing the learning curve for designers already accustomed to writing scripts for Vitis HLS. 
The main feature of OptDSL is the bracket notation that parameterizes an optimization directive with multiple choices.
The overall design space is the Cartesian product of the choices for each parameterized directive.

\subsubsection{Vendor-specific Concrete Design Generation}\label{sec:frontend-extensibility}

While abstract designs can be vendor-agnostic, concrete designs are vendor-specific. I.e., different vendor tools have different HLS syntax and directive formats;
therefore, during the lowering process, the frontend needs vendor-specific logic to target different tool flows, as depicted in Fig.~\ref{fig:overview} stage 1.
\ourwork{} currently provides support for AMD/Xilinx and Intel flows, while other vendors can be easily supported.

The OptDSL file is provided within the abstract design as a file named \texttt{opt\_\-template.tcl}. To lower the abstract design for AMD/Xilinx tools, we generate \texttt{opt.tcl}, a version of \texttt{opt\_\-template.tcl} with bracketed parameters replaced with different concrete values for each design point.
Once these bracketed parameters are substituted, the OptDSL script becomes a valid Tcl script that can be used directly with Vitis HLS.

To support other vendors, the frontend can parse the OptDSL file and identify the specific optimization directives used within it together with their parametrizations.
If the provided OptDSL is not sufficient to describe a desired DSE, \ourwork{} provides the necessary infrastructure to allow users to specify their own entirely custom frontend as Python code, as long as it conforms to the specified API interface (to be discussed in Sec.~\ref{sec:api}). For instance, a new frontend pass can easily be introduced to parameterize constants in the HLS source code itself: simply copy the existing OptDSL frontend and modify the templating logic and syntax to work with files other than \texttt{opt\_template.tcl}.

\subsubsection{Design Space Sampling}
\label{sec:dss}

The design space created by the parameterized optimization directives may be extremely large for even a single design, growing exponentially with the addition of each directive. Therefore, it is almost impossible to enumerate every possible design point in the specified design space and execute synthesis and implementation.

\ourwork{} natively supports random sampling of design points from the Cartesian product of all combinations of optimization directives. Users can specify the number of sampled design points, trading off design space coverage for dataset build time and storage.

\textit{Unlike design space exploration}, \ourwork{}'s enumeration and random sampling approaches are not guided or optimization-driven. The focus is solely on collecting a wide range of designs, including suboptimal designs, which are important for building ML datasets and training ML models that can interpolate to as many unseen designs during evaluation and deployment, not just optimal designs.

In the future, \ourwork{} can be extended to support user-customizable heuristics for selecting design points, utilizing expert knowledge to determine which combinations of optimizations are more useful to sample from and which combinations may result in invalid or redundant designs. 
For example, the sampling stage can be combined with active learning to determine meaningful design points to be synthesized.

\subsection{Stage 2: Design Synthesis}\label{sec:design-synthesis}

The second stage of \ourwork{} synthesizes and implements each concrete HLS design, a process we collectively refer to as the design synthesis. This stage also has an entry point for user input---vendor-specific concrete designs can be provided directly at this point without going through design space expansion. This is useful for easy integration of third-party HLS designs where parametrization of the design space may be difficult or unnecessary.

Design synthesis is broken down into two steps: (1) \texttt{HLSSynth}, where an HLS design is synthesized to RTL code, and (2) \texttt{HLSImpl}, where the resulting RTL code is implemented, resulting in a fully placed-and-routed design. For AMD/Xilinx designs, Vitis HLS is used for \texttt{HLSSynth} and Vivado for \texttt{HLSImpl}. However, any vendor tool can easily be integrated into the \ourwork{} framework, for example, Yosys~\cite{yosys} or Intel HLS (to be demonstrated in Sec.~\ref{sec:case-study-1}), by providing Python code for the desired \texttt{ToolFlow} subclasses.

\subsection{Stage 3: Data Extraction and Aggregation}

Once all the frontends and tool flows have been executed on a pool of designs, relevant design data must be extracted and aggregated into structured formats.
\ourwork{} provides \texttt{DataAggregator} classes to package HLS synthesis data (estimated latency, resource usage), post-implementation data (timing, resource, and power data), tool execution metadata (version, runtime), and build artifacts (LLVM IR, IP blocks) into shareable datasets.

Furthermore, as in stage 2, users may want to provide input directly at this stage, e.g., when integrating pre-generated data from prior works, where the build process is not reproducible and thus an earlier entry point cannot be used. Therefore, \ourwork{} provides an entry point to the data aggregation stage. This entry point can accept fully synthesized and implemented designs, from which \ourwork's built-in data aggregators can extract the relevant data, or pre-generated metrics in whatever form is available, which can be used with a custom \texttt{DataAggregator} subclass to adapt such metrics into \ourwork's standard output format.

\section{Implementation and Usage}\label{sec:implementation-and-usage}
\subsection{Vendor Agnostic User API}\label{sec:api}

\begin{figure}[t]
    \centering
    \scriptsize
\begin{lstlisting}[
basicstyle=\scriptsize\selectfont\ttfamily, 
language=Python
]
datasets: DesignDatasetCollection = {
    "polybench_xilinx": dataset_polybench_builder(WORK_DIR),
    "machsuite_xilinx": dataset_machsuite_builder(WORK_DIR),
    "chstone_xilinx": dataset_chstone_builder(WORK_DIR),
}

opt_dsl = OptDSLFrontend(WORK_DIR, random_sample=True,
                         random_sample_num=N_RANDOM_SAMPLES)
hls_synth = VitisHLSSynthFlow()
hls_impl = VitisHLSImplFlow()
hls_impl_report = VitisHLSImplReportFlow()

datasets_post_frontend = opt_dsl.execute_datasets_parallel(
    datasets, n_jobs=N_JOBS)
datasets_post_synth = hls_synth.execute_datasets_parallel(
    datasets_post_frontend, n_jobs=N_JOBS)
datasets_post_hls_impl = hls_impl.execute_datasets_parallel(
    datasets_post_synth, n_jobs=N_JOBS)
hls_impl_report.execute_datasets_parallel(
    datasets_post_hls_impl, n_jobs=N_JOBS)
\end{lstlisting}
    \vspace{-2.0em}
    \caption{Example usage of the \ourwork{} framework.}
    \label{listing:usage}
\end{figure}

\ourwork{} is implemented as a Python library and provides a simple user API that allows the framework configuration to be expressed easily as a short Python script (while still allowing for full Python programming if complex configuration is desired).

\begin{table}[th]
    \centering
    \small
    \caption{The \ourwork{} User API.}
    \vspace{-1.0em}
    \resizebox{\columnwidth}{!}{
        \renewcommand{\arraystretch}{0.8}
        \begin{tabular}{l|c}
            \toprule
            \textbf{API Functions}                    & \textbf{Description}                                   \\ \midrule
            \texttt{class Design}                     & Single HLS design                                      \\
            \texttt{class Dataset}                    & Multiple HLS designs                                   \\
            \midrule
            \texttt{class Flow(ABC)}              & Abstract class for arbitrary design flow   \\
            \texttt{Flow.execute(design)}   & Execute a flow on one design                       \\
            \texttt{Flow.execute\_datasets\_parallel(design)}              & Execute a flow on many designs                      \\
            \midrule
            \texttt{class Frontend(Flow)}              & Abstract class for frontend design expansion   \\
            \texttt{class OptDSLFrontend(Frontend)}         & Opt DSL frontend for Xilinx HLS designs                \\

            \midrule
            \texttt{class ToolFlow(Flow)}              & Abstract class for EDA tool                      \\
            \texttt{class VitisHLSSynthFlow(ToolFlow)}      & Run Vitis HLS synthesis                         \\
            \texttt{class VitisHLSImplFlow(ToolFlow)}       & Run Vivado implementation (via Vitis HLS) \\
            \texttt{class VitisHLSImplReportFlow(ToolFlow)} & Run Vivado reporting                      \\

            \bottomrule
        \end{tabular}
    }
    \vspace{0.5em}
    \label{tab:hlsdataset-api}
\end{table}

An example is shown in Fig.~\ref{listing:usage}. The source HLS designs are located and copied to the desired work directory, and the \texttt{OptDSLFrontend} is invoked to sample 10 random design points from each design. The \texttt{VitisHLSSynthFlow} and \texttt{VitisHLSImplFlow} are then be invoked to synthesize and implement each design point, followed by data aggregation using the \texttt{VitisHLSImplReportFlow} to gather data from each implemented design in a standardized format. A full list of the available APIs is available in Table~\ref{tab:hlsdataset-api}.

The API also includes abstract base classes (\texttt{ABC}s) that users can subclass to implement their own frontends and tool flows for \ourwork, for instance, to support another vendor's HLS tools. 
\ourwork{} abstracts away the complexities of integrating custom user subclasses into the overall dataset generation process, including the use of fine-grained parallelism (to be discussed in Sec.~\ref{sec:parallelism}).

\subsection{Directory Structure}

\begin{figure}

    \centering
    \vstretch{0.95}{\includegraphics[width=\linewidth]{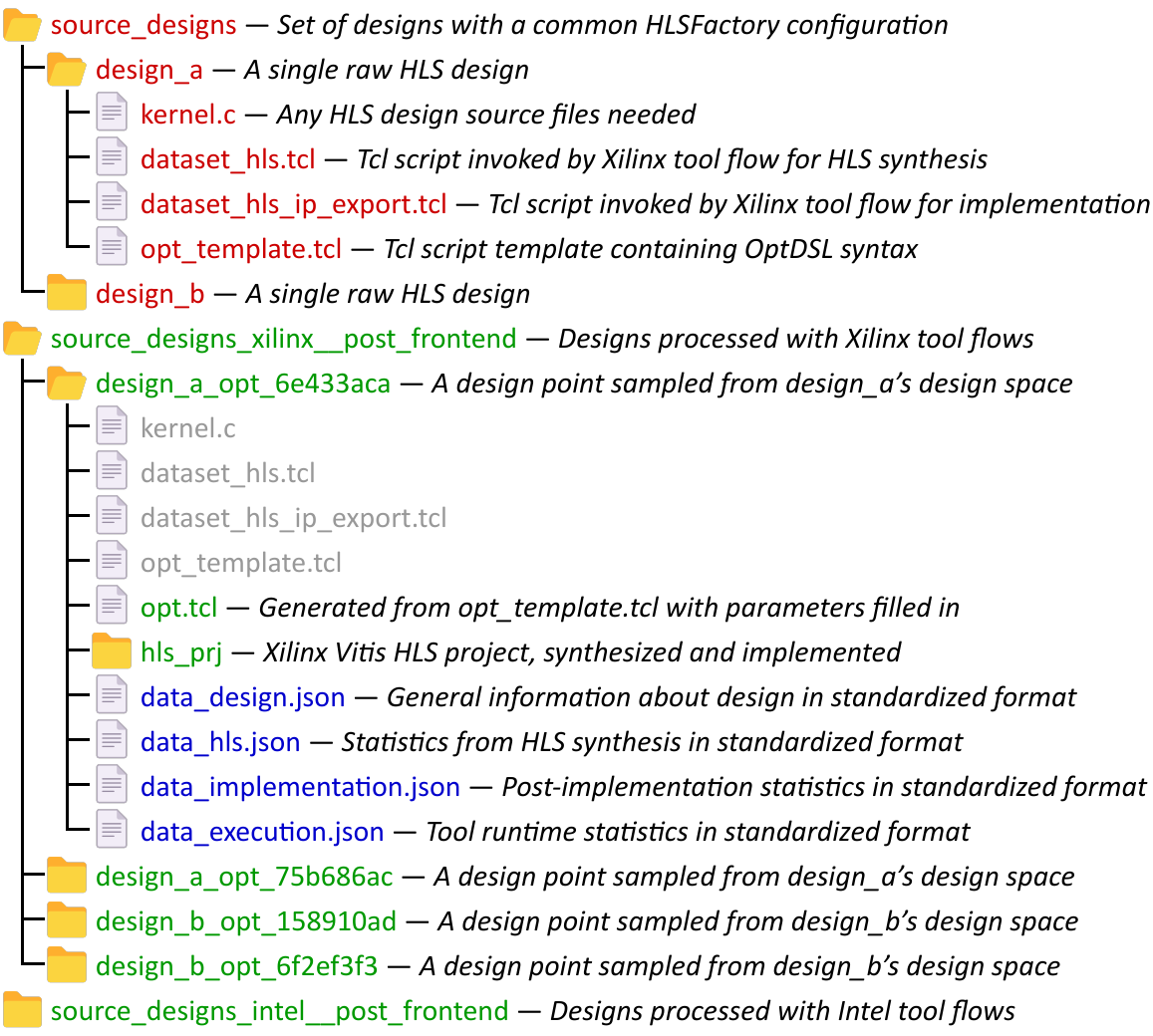}}
    \vspace{-2.0em}
    \caption{The directory structure that \ourwork{} uses. Red are input files; green are the intermediate design points; blue are output files.
    }
    \label{fig:filesystem-structure}

\end{figure}

Fig.~\ref{fig:filesystem-structure} depicts a simple example of the directory structure accepted as input and produced as output of the \ourwork{} workflow. As described throughout Sec.~\ref{sec:framework}, we first sample the design space for each source abstract design and then run tool flows and data aggregation on the sampled concrete designs. The figure presents the directory structure for the inputs to this process: an abstract design specified in terms of HLS kernel code, an \texttt{opt\_template.tcl} file to be used by the OptDSL frontend (described in Sec.~\ref{sec:opt-dsl-frontend}), and auxiliary scripts for the AMD/Xilinx tool flows.

During dataset generation, each abstract design is enumerated into multiple concrete designs, shown in the figure under the newly generated directory \texttt{source\_designs\_xilinx\_\_post\_frontend}. Each concrete design is identified by the concatenation of the name of the original abstract design and a unique hash determined by the combination of optimization directives chosen for that design. This unique combination of optimization directives is generated as the concrete design's \texttt{opt.tcl} file.

Tool flows and data aggregation run directly within these concrete design directories. After HLS projects are created, synthesized, and implemented (within the \texttt{hls\_prj} directory, as depicted), the data aggregation stage collects information from these projects into standardized JSON-formatted files. These JSON files are stored alongside the HLS project directory within each concrete design, making it clear exactly which combination of optimization directives were used to generate the data.

\subsection{Parallel Build Backend}\label{sec:parallelism}
To build datasets with hundreds and thousands of data points, an efficient backend is needed to dispatch and execute multiple frontend and tool flows in parallel. In the case of HLS, the bottleneck of constructing such datasets is the runtime of the vendor tools themselves. The runtime for synthesizing an HLS design can range from minutes to hours. We may also want to run trial FPGA implementation flows, which can take hours.

To address these needs, every frontend and tool flow component is automatically augmented in a fine-grained parallel build backend based on multiprocessing. Since all frontend and tool flows are based on the abstract base class, we can easily provide this facility to the user. We take advantage of Python's \texttt{multiprocessing}. We also provide the option to pin each task to its own dedicated CPU core.
This approach appears to be a good default to distribute design build workloads on many-core systems.

We also provide a way for users to pool parallelism across dataset collections rather than a single dataset. Users are able to describe a collection of datasets, each with their own set of designs. Instead of dispatching each dataset's build workloads in its own parallel pool (i.e., naive parallelism), we aggregate all designs into a single parallel pool (i.e., fine-grained parallelism). This feature is automatic for every frontend and tool flow and transparent to the end user.

\section{Evaluations}\label{sec:results}

We evaluate our work through a series of seven case studies which demonstrate \ourwork{}'s multifunctionality and ease of use.

\subsection{Case Study 1: ML Prediction of Post-Implementation QoR}
\label{sec:qor}

\begin{figure}[b]
    \centering
    \includegraphics[width=\linewidth]{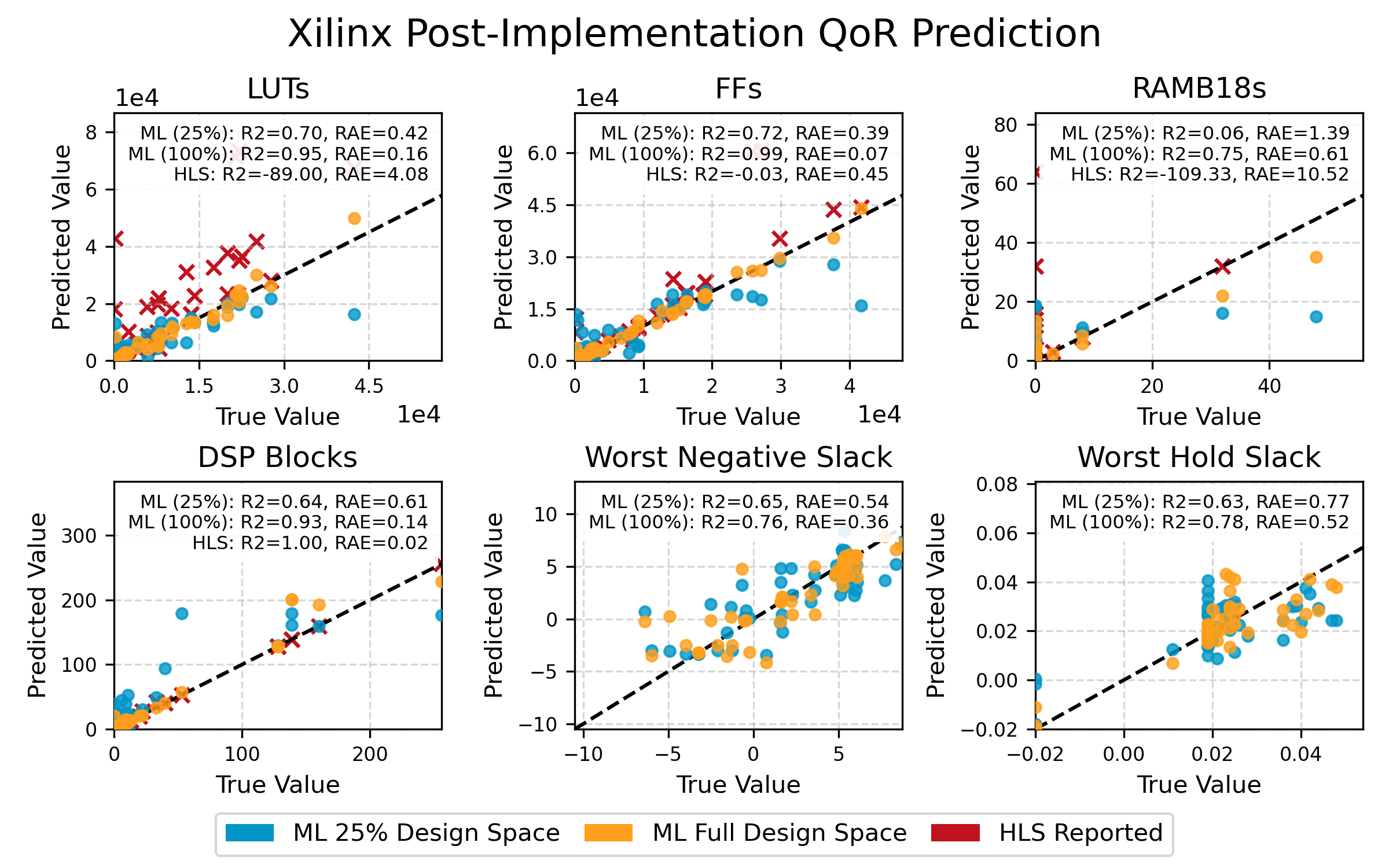}
    \vspace{-2.0em}
    \caption{True-vs-predicted plots for the HLS-based ML QoR model. Test values are shown for models trained on the complete and partial subset of the training design space. ``RAE": Relative Absolute Error ($|\hat{y} - y| / |y - \bar{y}|$), ``R2": Coefficient of Determination}
    \label{fig:hls-qor}
\end{figure}

HLS vendor tools provide resource usage estimates (e.g., \#LUTs, \#FFs, \#DSPs, \#BRAMs) and timing information (e.g., II violations, clock speed) for designs based on scheduling and binding results. However, HLS-estimated results often deviate significantly from post-implementation resource usage and may not correlate well with critical timing metrics (e.g., worst negative slack and worst hold slack). Previous works, such as S. Dai et al. \cite{quickest}, address this issue by using ML-based models to predict post-implementation quality-of-results (QoR) metrics based on HLS-reported metrics.

We demonstrate that \ourwork{} can replicate the approach used by S. Dai et al. \cite{quickest} to build ML models for post-implementation QoR predictions targeting Vitis HLS and Vivado. We use \ourwork{} built-in Polybench, MachSuite, and CHStone design datasets which provide $n=29$ base designs; using the the OptDSL frontend, design space expansion is performed resulting in $n=257$ final designs. \ourwork{}'s APIs are also used run tool synthesis and implementation as well as bundle the HLS post-implementation data into a tabular dataset. A histogram-based gradient boosting regression model is then trained to predict post-implementation reported resources and timing metrics using HLS-reported resources, latency, clock speed, and arithmetic/logic operation counts as model inputs. We train our model on an 80\%/20\% train-test split, as well as a 25\% subset of the training data to demonstrate the utility of design space expansion in improving ML model performance.

Our results, shown in Fig.~\ref{fig:hls-qor}, indicate that the $R^2$ value and mean relative error are better for the larger training set achieved through design space expansion. We highlight that generating more data points using \ourwork{}'s design space expansion will result in higher prediction accuracy, even when randomly sampling from the entire design space and including suboptimal, i.e. ``bad", designs (in terms of QoR metrics). For most resource prediction targets, our ML model also has a lower relative error than the HLS-reported values, showing improvement over the HLS tool itself. These results highlight the utility of HLSFactory applied to ML for EDA and the importance of design space expansion, even with a smaller sample size, for robust ML dataset construction and model training.

\subsection{Case Study 2: Design Space Coverage}
\label{sec:dsc}

We evaluate how the use of design space expansion in \ourwork{} quantitatively and qualitatively improves the overall design space of generated datasets in terms of latency (HLS-reported) and resource usage (post-implementation). In the context of ML, improved design space coverage for these metrics is important for robust model training on downstream tasks, such as ML-based QoR prediction. Thus we perform a case study comparing metrics of the base designs in Polybench, MachSuite, and CHStone ($n=29$) with the designs sampled from them ($n=257$); this is the same dataset used in \S\ref{sec:qor}.

\begin{figure}[t]
    \centering
    \includegraphics[width=\linewidth]{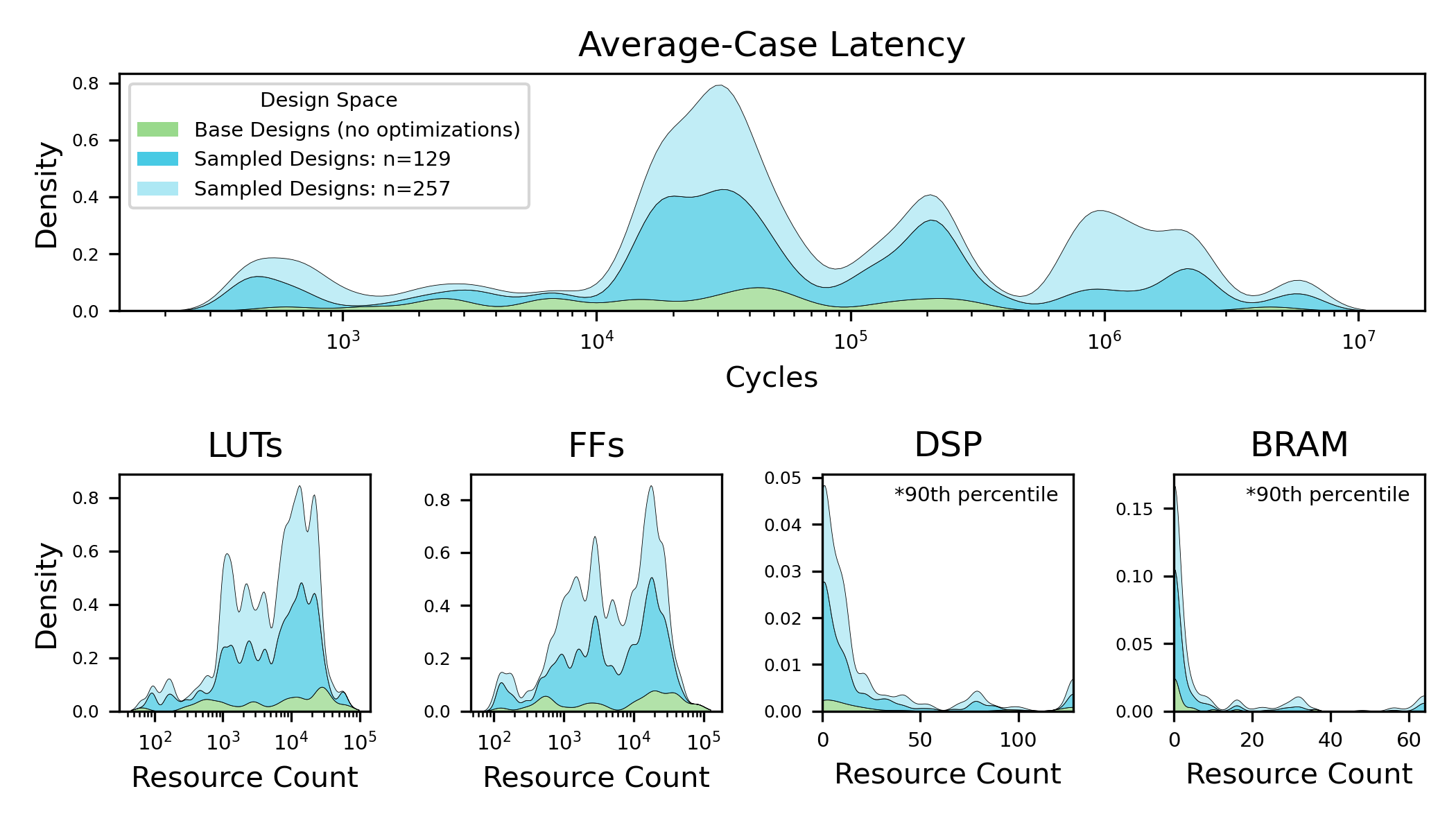}
    \vspace{-2.5em}
    \caption{Effect of design sampling to cover more design space.  Sampled designs cover a wider range of metrics than base designs with no optimizations. Latency is HLS estimated; resources are post-implementation. Note that these are \textit{stacked} density plots to show the effect of cumulative design sampling.}
    \label{fig:metrics-design-level}
\end{figure}

We start with a quantitative evaluation. Fig.~\ref{fig:metrics-design-level} illustrates the cumulative distributions of these metrics as a stacked histogram representing only base designs ($n=29$), half the sampled designs ($n=129$), and all the sampled designs ($n=257$). We highlight that the sampled designs cover a wider range of average-case latency, LUT usage, and FF usage, with denser coverage as $n$ increases. In the case of DSP and BRAM usage, most base designs use none of these resources while sampled designs do.

\begin{figure}[ht]
    \centering
    \includegraphics[width=\linewidth]{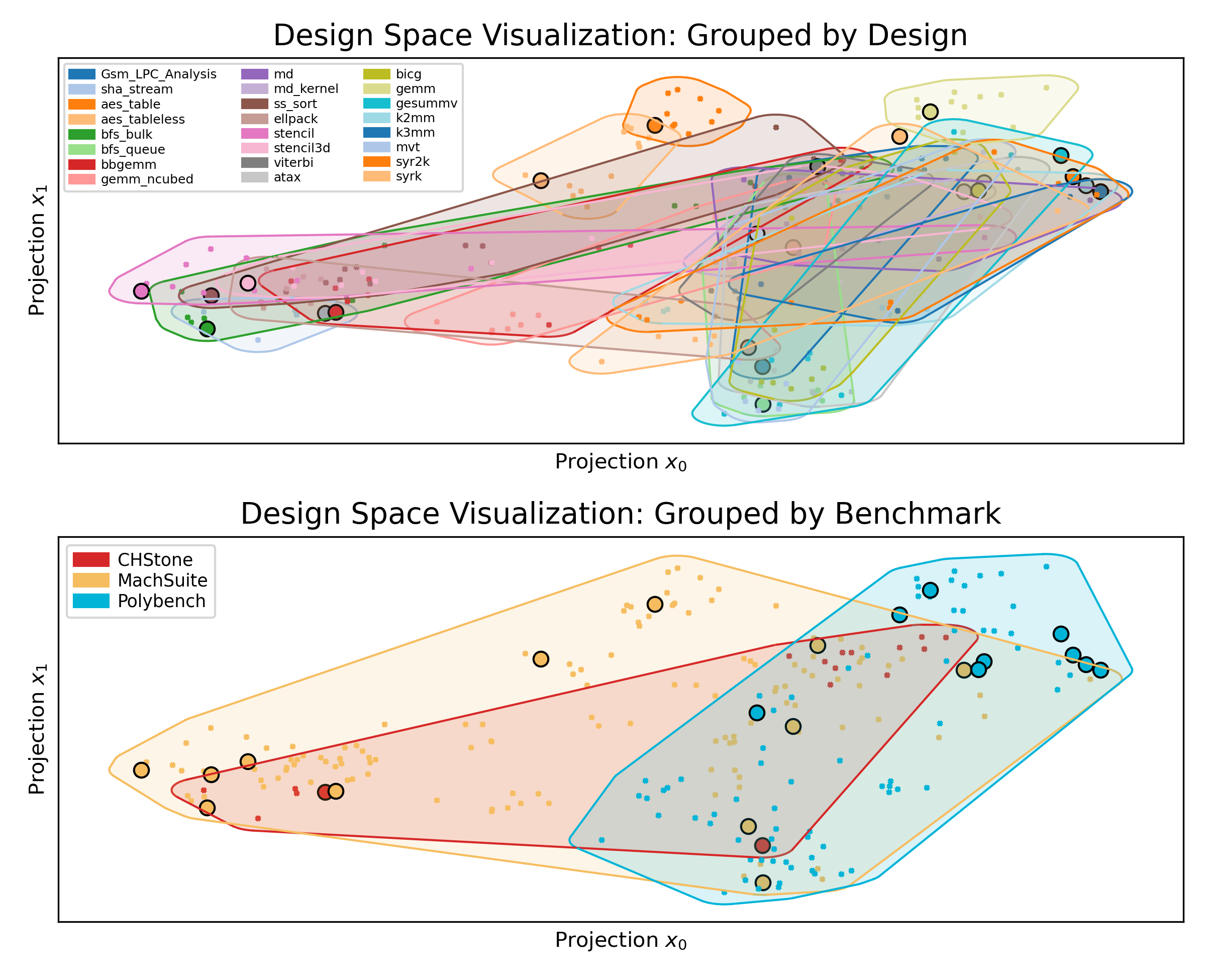}
    \vspace{-2.5em}
    \caption{Embedding of sampled designs across selected benchmarks. Base designs without optimizations are emphasized. Design points and locations are the same between both panels; they are only colored and grouped differently.}
    \label{fig:ds-all}
\end{figure}

We then illustrate the qualitative coverage of the design space in Fig.~\ref{fig:ds-all}. This space is the 2-D embedding space of HLS-reported and post-implementation metrics generated using a PacMAP~\cite{pacmap} dimensional reduction.
Each of the base designs is depicted as large emphasized points within this embedding space; sampled designs from the same base design (top panel) or the same benchmark (bottom panel) have matching colors. The convex hulls around same-colored points show the portion of the embedding space covered by design space expansion from each base design or benchmark. This clearly shows that sampling from the expanded design space results in non-overlapping coverage that otherwise would not appear in the final dataset.

\begin{figure}[t]
    \centering
    \includegraphics[width=0.95\linewidth]{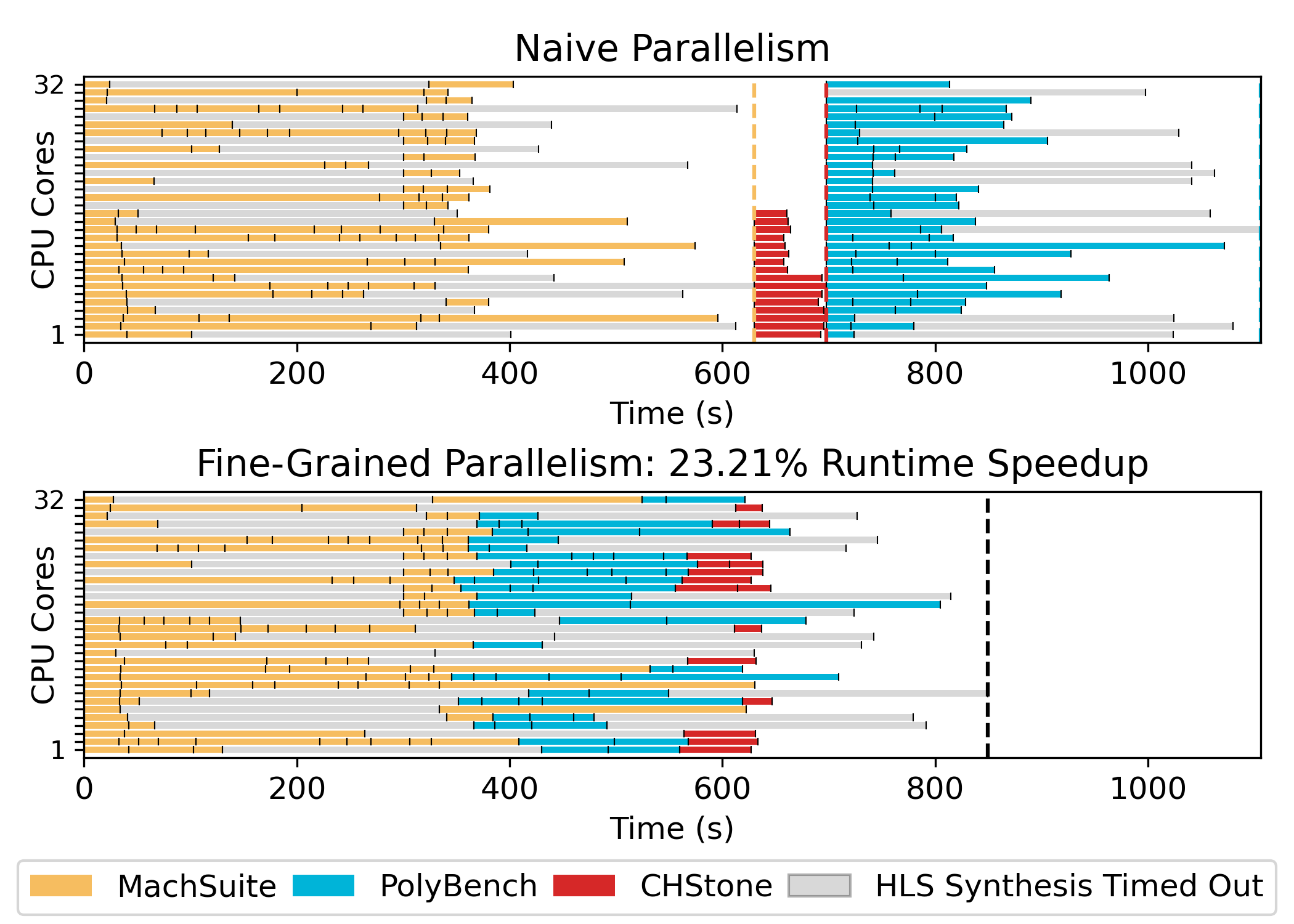}
    \vspace{-1em}
    \caption{Parallel execution of Vitis HLS synthesis. Top panel shows core utilization over time with naive parallelism across datasets;
    bottom panel shows our fine-grained design parallelism across datasets.
    }
    \label{fig:parallel-build-results}
\end{figure}

\subsection{Case Study 3: Speedup of Fine-Grained Design Parallelism}

We evaluate our fine-grained parallelism strategy described in Sec.~\ref{sec:parallelism} using a case study synthesizing designs sampled from Polybench, MachSuite, and CHStone using Vitis HLS across 32 CPU cores.

Results are shown in Fig.~\ref{fig:parallel-build-results}, showing that fine-grained parallelism achieves more than 20\% speed up compared with the naive parallelism approach. 
Such fine-grained parallelism is especially beneficial given the user-specified timeout threshold (annotated as gray bars).

\subsection{Case Study 4: Targeting Different Vendors}
\label{sec:case-study-1}
To demonstrate the extensibility of the first stage of \ourwork{}, we show how to add support for Intel's i++ HLS flow.

As described in Sec.~\ref{sec:frontend-extensibility}, \ourwork{} includes an OptDSL parser that recognizes Vitis HLS optimization directives in \texttt{opt\_template.tcl}, such as the \texttt{set\_\-directive\_\-unroll} and \texttt{set\_\-directive\_\-array\_\-partition} commands. We can therefore build our Intel-lowering frontend on top of this functionality.

Because i++ does not support specifying optimization directives in a separate file, our frontend instead transforms the HLS source code directly to add i++-compatible versions of each directive parsed from the \texttt{opt\_template.tcl} file.

While our frontend can often generate exact equivalents for the specified directives, in some cases, i++ has no exact equivalent for a particular directive used by Vitis HLS, such as \texttt{array\_\-partition} directives. In these cases, we substitute similar directives---in this case, a combination of Intel directives \texttt{hls\_numbanks} and \texttt{hls\_bankwidth} that achieve a similar memory partitioning result.

Since \ourwork{} is agnostic to the specific directives being used and does not correlate specific AMD/Xilinx concrete designs with specific Intel concrete designs, directives need not match one-to-one. There is no impact on correctness; substituting similar directives still improves the diversity of the dataset.

\begin{figure}[t]
    \centering
    \includegraphics[width=\linewidth]{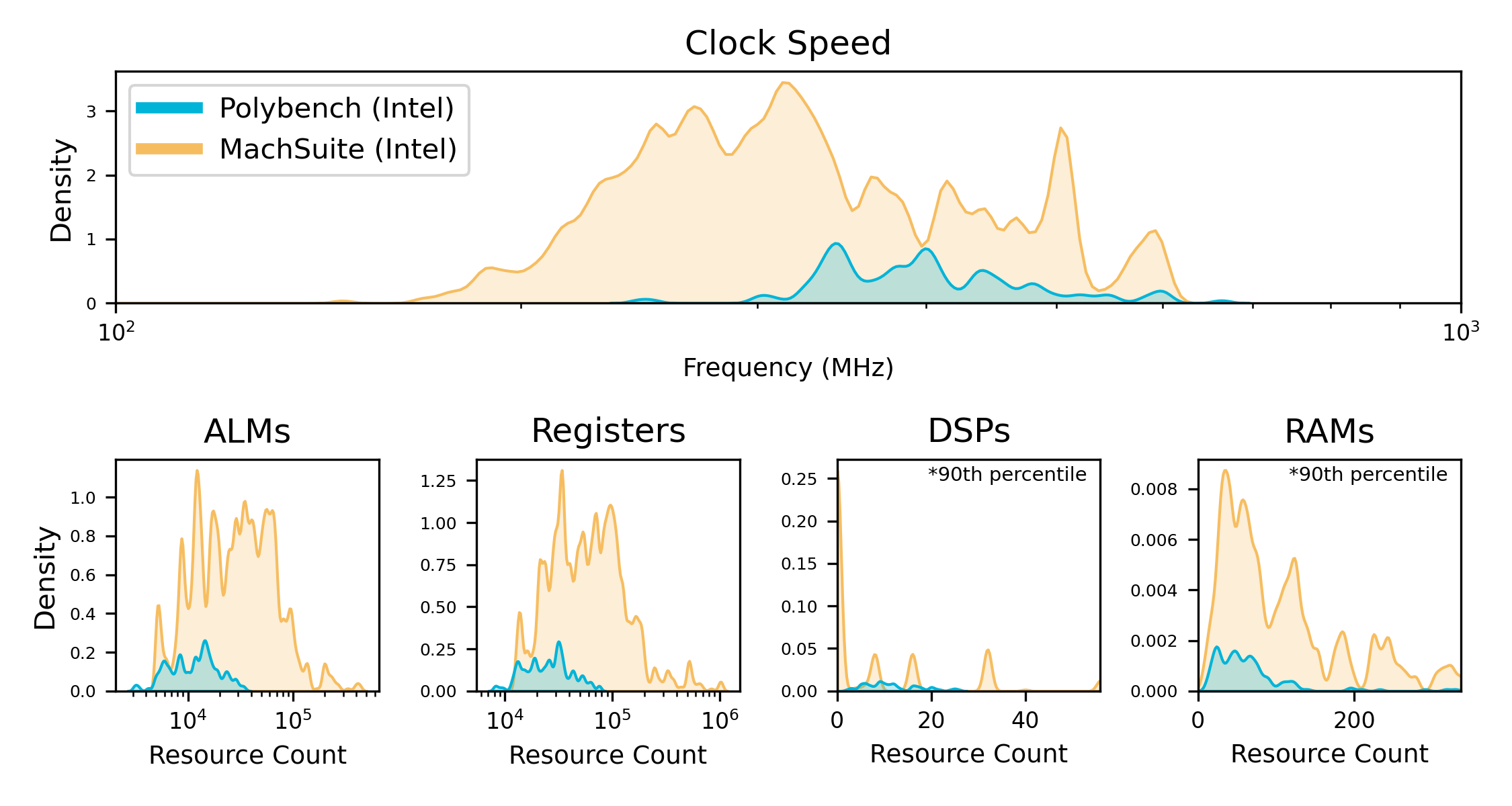}
    \vspace{-2.5em}
    \caption{Distribution of post-implementation metrics for PolyBench and MachSuite designs ($n=1340$) using Intel's HLS flow.}
    \label{fig:intel-dist}
\end{figure}

In total, our end-to-end Intel flow extends the \ourwork{} user APIs in Table~\ref{tab:hlsdataset-api} with three Intel equivalents:
\texttt{Opt\-DSL\-Frontend\-Intel\-(Frontend)} as described above, \texttt{Intel\-HLS\-Synth\-Flow\-(Tool\-Flow)} to invoke i++ for HLS, and \texttt{Intel\-Quartus\-Impl\-Flow\-(Tool\-Flow)} to invoke Quartus for implementation. We run this flow on designs sampled from PolyBench and MachSuite and plot the resulting metrics in Fig.~\ref{fig:intel-dist}. Intel's HLS tool does not report overall latency estimates, but it optimizes each kernel's throughput by maximizing clock speed, which we use as a proxy for performance.

\subsection{Case Study 5: Adding Auxiliary Design Collections}
\label{sec:case-study-2}

Third-party researchers may have existing, synthesizable, vendor-specific HLS designs to integrate into \ourwork, but they may not want or need to create an OptDSL specification for them.
For instance, the authors of LightningSim~\cite{lightningsim} collect 33 synthesizable open-source designs for AMD/Xilinx Vitis HLS to evaluate their simulation tool, including designs from AMD/Xilinx sample code repositories~\cite{xilinx2021basic,xilinx2022vitis}, algorithm implementations from Kastner \textit{et al.}'s \textit{Parallel Programming for FPGAs}~\cite{kastner2018parallel}, and graph neural network implementations from FlowGNN~\cite{flowgnn}. These designs are all provided in a standard format, each having a Tcl script \texttt{setup.tcl} to set up a Vitis HLS project for synthesis.

Using the entry point at the design synthesis stage, one graduate student was able to integrate all of these designs into \ourwork{} in less than one hour. To match the input directory structure in Fig.~\ref{fig:filesystem-structure}, we only needed to copy \texttt{setup.tcl} to \texttt{dataset\_\-hls.tcl} with \texttt{csynth\_\-design} appended (\ourwork's \texttt{Vitis\-HLS\-Synth\-Flow} expects it to setup the project \textit{and} run synthesis) and add a four-line script \texttt{dataset\_\-hls\_\-ip\_\-export.tcl} to invoke implementation from Vitis HLS.
Since we used the entry point after design space expansion, these were concrete designs, not abstract designs, so no \texttt{opt\_template.tcl} was required.

Many other works~\cite{dgnnbooster,masknet,skynet,edgemoe} were also easily integrated with \ourwork{} in a similar fashion; the code is available online.

\subsection{Case Study 6: Integrating Released Data from Other Works}
\label{sec:case-study-3}

\begin{figure}[t]
    \vspace{-0.5em}
    \centering
    \includegraphics[width=\linewidth]{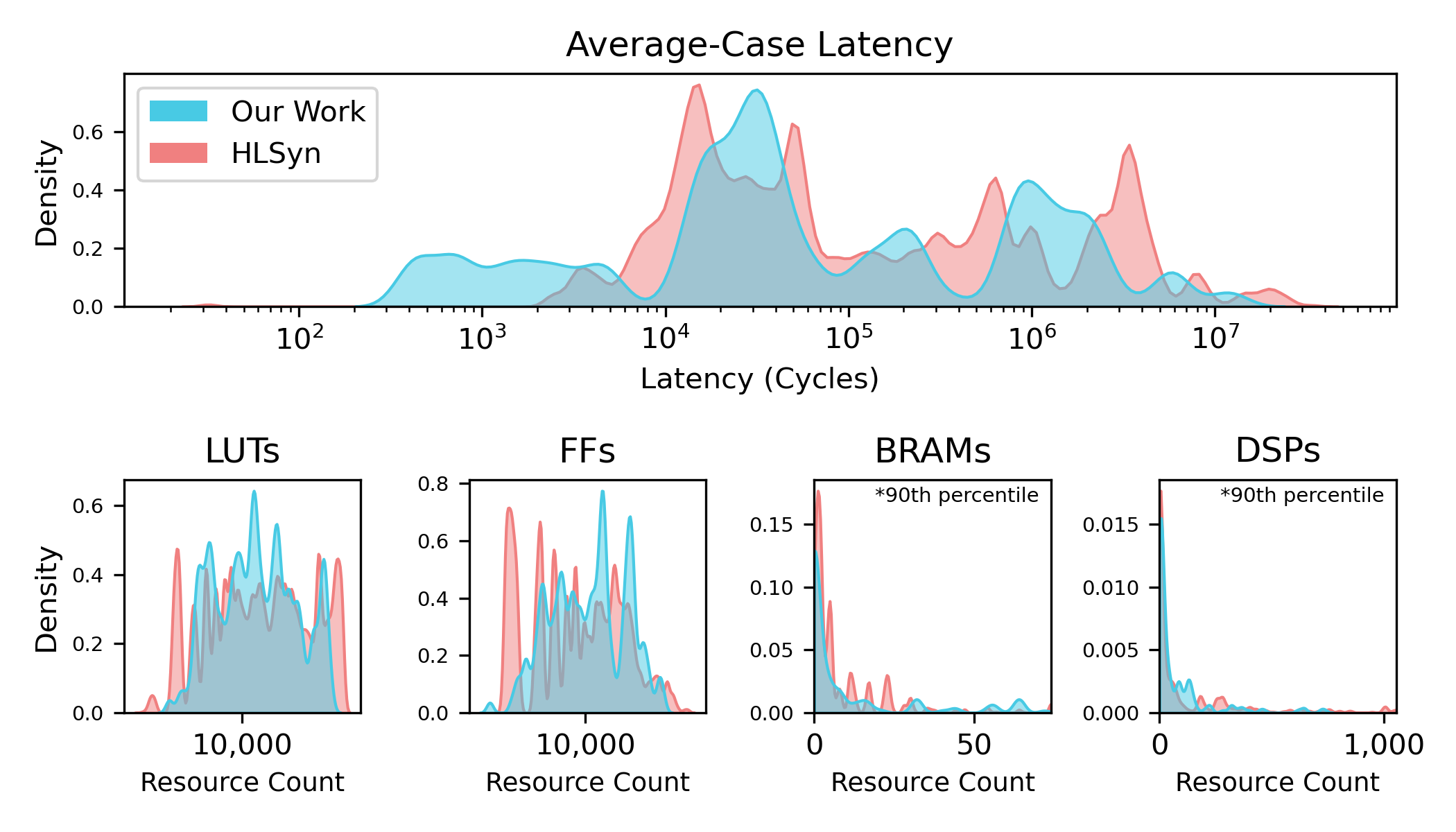}
    \vspace{-2.5em}
    \caption{Distribution of HLS-estimated metrics from selected \ourwork{} benchmarks (PolyBench, CHStone, and MachSuite; $n=167$) vs.\ HLSyn ($n=3371$).}
    \label{fig:hlsyn}
\end{figure}

We may still want to incorporate previously published data have published to build a more comprehensive HLS dataset. \ourwork{}'s data aggregation step provides an entry point to incorporate external data sources into our dataset with ease.

We illustrate how \ourwork{} can integrate pre-generated data from prior works---in this case, HLSyn~\cite{chang2023dr}. HLSyn provides both the source code (with places to template optimization directives) of their selected kernels, as well as associated metrics for HLS-reported resource usage and latency for sampled designs.
We write a \texttt{DataAggregator} subclass to integrate this data into HLSFactory.

The results are illustrated in Fig.~\ref{fig:hlsyn}, showing the distributions of reported HLS metrics sourced from the listed valid designs of HLSyn and a small sampled subset of designs from our base PolyBench, CHStone, and Machsuite datasets.

The HLSyn flow is built on top of AutoDSE \cite{autodse, autodse_software} and the Merlin compiler \cite{merlin-chapter, merlin}, both of which are open-source software tools aimed at optimized design space exploration (DSE) and source-to-source translation. These tools suggest future work to integrate AutoDSE and the Merlin compiler as custom flows in \ourwork{}, allowing designs to be built from the design space specifications defined in AutoDSE and synthesized with the Merlin compiler.

\begin{figure}[t]
    \centering
    \includegraphics[width=\linewidth]{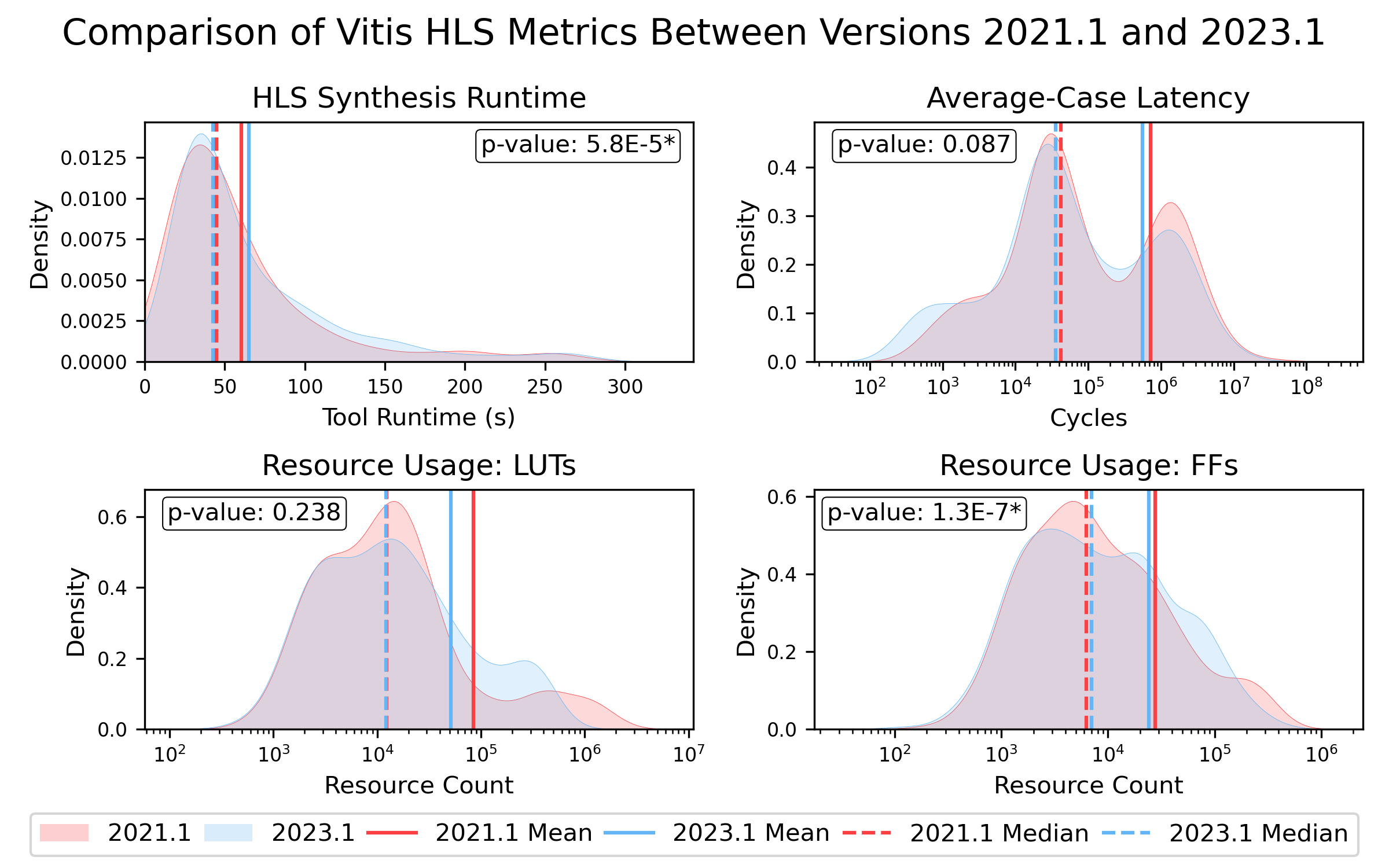}
    \vspace{-2.0em}
    \caption{Distribution of HLS tool metrics from two versions of Vitis HLS.}
    \label{fig:hls-regression}
\end{figure}

\subsection{Case Study 7: Regression Benchmarking HLS Synthesis Tools}
\label{sec:reg}

New versions of HLS vendor tools are periodically released and improve both the tool performance (e.g., faster synthesis) and the QoR of synthesized designs (e.g., less resource usage). However, quantifying such improvements across different tool versions is difficult without a way to benchmark a wide range of designs, similar to the regression testing used in traditional software development.

We demonstrate that \ourwork{} streamlines regression testing on HLS tools. We compare Vitis HLS versions 2021.1 and 2023.1 using designs sampled from Polybench, Machsuite, and CHStone (with 16 samples per base design). We collect paired samples by synthesizing the same design with both tool versions. 

This experiment was set up in a fully self-contained Python script and \ourwork{} enabled this initial study to be completed by one graduate student in three hours.

The results are shown in Fig.~\ref{fig:hls-regression}. We show distributions of the tool runtime, HLS-estimated latency, LUT usage, and FF usage across tool versions. We also report the $p$-value for a paired two-tailed Wilcoxon signed-rank test~\cite[p.~350]{conover1999practical}
and indicate cases with a $p$-value less than $\alpha=0.05$ with an asterisk, indicating a statistically significant difference. Note that for certain metrics, the mean and median shift in opposite directions between tool versions.

\section{Conclusion}\label{sec:conclusion}

\ourwork{} brings a much-needed principled approach to generating datasets of HLS designs. Our case studies show a small sample of what can be done when a flexible, reproducible way to generate data from HLS designs is available. 
We demonstrate that there is substantial untapped potential for future research into how ML can be applied to HLS.

We also consider directions for future extensions of \ourwork. Our framework currently has no support for collecting post-simulation metrics like vector-based power analysis or simulated latency. Introducing simulation to \ourwork, particularly for designs where only a high-level C testbench is available rather than an RTL testbench, is a valuable direction for future work.

We hope that, through open-source, this work invites the research community to collaborate and contribute more designs and tool flows and accelerate ML research for EDA applications.

\pagebreak

\section*{Acknowledgements}
This research was supported in part by  National Science Foundation (NSF) Grant \#2326894, NVIDIA Applied Research Accelerator Program Grant, and the Texas Advanced Computing Center (TACC). Any opinions, findings, conclusions, or recommendations are those of the authors and not of the funding agencies. We also thank Georgia Tech Research Institute for direct funding of selected authors.

\bibliographystyle{ACM-Reference-Format}
\bibliography{refs}


\begin{thebibliography}{38}


\ifx \showCODEN    \undefined \def \showCODEN     #1{\unskip}     \fi
\ifx \showDOI      \undefined \def \showDOI       #1{#1}\fi
\ifx \showISBNx    \undefined \def \showISBNx     #1{\unskip}     \fi
\ifx \showISBNxiii \undefined \def \showISBNxiii  #1{\unskip}     \fi
\ifx \showISSN     \undefined \def \showISSN      #1{\unskip}     \fi
\ifx \showLCCN     \undefined \def \showLCCN      #1{\unskip}     \fi
\ifx \shownote     \undefined \def \shownote      #1{#1}          \fi
\ifx \showarticletitle \undefined \def \showarticletitle #1{#1}   \fi
\ifx \showURL      \undefined \def \showURL       {\relax}        \fi
\providecommand\bibfield[2]{#2}
\providecommand\bibinfo[2]{#2}
\providecommand\natexlab[1]{#1}
\providecommand\showeprint[2][]{arXiv:#2}

\bibitem[pol({[n.\,d.]})]%
        {polybench}
 \bibinfo{year}{[n.\,d.]}\natexlab{}.
\newblock \bibinfo{booktitle}{\emph{{{PolyBench}}}}.
\newblock
\urldef\tempurl%
\url{https://web.cse.ohio-state.edu/~pouchet.2/software/polybench/}
\showURL{%
\tempurl}


\bibitem[aut({[n.\,d.]})]%
        {autodse_software}
UCLA VAST Lab \bibinfo{year}{[n.\,d.]}\natexlab{}.
\newblock \bibinfo{booktitle}{\emph{{{UCLA-VAST}}/{{AutoDSE}}}}.
\newblock UCLA VAST Lab.
\newblock
\urldef\tempurl%
\url{https://github.com/UCLA-VAST/AutoDSE}
\showURL{%
\tempurl}


\bibitem[mer({[n.\,d.]})]%
        {merlin}
{Xilinx} \bibinfo{year}{[n.\,d.]}\natexlab{}.
\newblock \bibinfo{booktitle}{\emph{Xilinx/Merlin-Compiler}}.
\newblock {Xilinx}.
\newblock
\urldef\tempurl%
\url{https://github.com/Xilinx/merlin-compiler}
\showURL{%
\tempurl}


\bibitem[{AMD/Xilinx}(2021)]%
        {xilinx2021basic}
\bibfield{author}{\bibinfo{person}{{AMD/Xilinx}}.} \bibinfo{year}{2021}\natexlab{}.
\newblock \bibinfo{title}{Basic Examples for {{Vitis HLS}}}.
\newblock \bibinfo{howpublished}{GitHub}.
\newblock


\bibitem[{AMD/Xilinx}(2022)]%
        {xilinx2022vitis}
\bibfield{author}{\bibinfo{person}{{AMD/Xilinx}}.} \bibinfo{year}{2022}\natexlab{}.
\newblock \bibinfo{title}{Vitis Accel Examples' Repository}.
\newblock \bibinfo{howpublished}{GitHub}.
\newblock


\bibitem[Bai et~al\mbox{.}(2023)]%
        {chang2023dr}
\bibfield{author}{\bibinfo{person}{Yunsheng Bai}, \bibinfo{person}{Atefeh Sohrabizadeh}, \bibinfo{person}{Zongyue Qin}, \bibinfo{person}{Ziniu Hu}, \bibinfo{person}{Yizhou Sun}, {and} \bibinfo{person}{Jason Cong}.} \bibinfo{year}{2023}\natexlab{}.
\newblock \showarticletitle{Towards a Comprehensive Benchmark for High-Level Synthesis Targeted to {{FPGAs}}}. In \bibinfo{booktitle}{\emph{Thirty-Seventh {{Conference}} on {{Neural Information Processing Systems Datasets}} and {{Benchmarks Track}}}}.
\newblock


\bibitem[Chen and Hao(2022)]%
        {masknet}
\bibfield{author}{\bibinfo{person}{Hanqiu Chen} {and} \bibinfo{person}{Cong Hao}.} \bibinfo{year}{2022}\natexlab{}.
\newblock \showarticletitle{Mask-{{Net}}: A Hardware-Efficient Object Detection Network with Masked Region Proposals}. In \bibinfo{booktitle}{\emph{2022 {{IEEE}} 33rd {{International Conference}} on {{Application-specific Systems}}, {{Architectures}} and {{Processors}} ({{ASAP}})}}. \bibinfo{publisher}{IEEE}, \bibinfo{address}{Gothenburg, Sweden}, \bibinfo{pages}{131--138}.
\newblock
\showISBNx{978-1-66548-308-7}
\urldef\tempurl%
\url{https://doi.org/10.1109/ASAP54787.2022.00030}
\showDOI{\tempurl}


\bibitem[Chen and Hao(2023)]%
        {dgnnbooster}
\bibfield{author}{\bibinfo{person}{Hanqiu Chen} {and} \bibinfo{person}{Cong Hao}.} \bibinfo{year}{2023}\natexlab{}.
\newblock \showarticletitle{{{DGNN-booster}}: A Generic {{FPGA}} Accelerator Framework for Dynamic Graph Neural Network Inference}. In \bibinfo{booktitle}{\emph{2023 {{IEEE}} 31st {{Annual International Symposium}} on {{Field-Programmable Custom Computing Machines}} ({{FCCM}})}}. \bibinfo{publisher}{IEEE}, \bibinfo{address}{Marina Del Rey, CA, USA}, \bibinfo{pages}{195--201}.
\newblock
\showISBNx{9798350312058}
\urldef\tempurl%
\url{https://doi.org/10.1109/FCCM57271.2023.00029}
\showDOI{\tempurl}


\bibitem[Chhabria et~al\mbox{.}(2021)]%
        {ml_eda1}
\bibfield{author}{\bibinfo{person}{Vidya~A. Chhabria}, \bibinfo{person}{Yanqing Zhang}, \bibinfo{person}{Haoxing Ren}, \bibinfo{person}{Ben Keller}, \bibinfo{person}{Brucek Khailany}, {and} \bibinfo{person}{Sachin~S. Sapatnekar}.} \bibinfo{year}{2021}\natexlab{}.
\newblock \showarticletitle{MAVIREC: ML-Aided Vectored IR-Drop Estimation and Classification}. In \bibinfo{booktitle}{\emph{2021 Design, Automation \& Test in Europe Conference \& Exhibition (DATE)}}.
\newblock


\bibitem[Cong et~al\mbox{.}({[n.\,d.]})]%
        {merlin-chapter}
\bibfield{author}{\bibinfo{person}{Jason Cong}, \bibinfo{person}{Muhuan Huang}, \bibinfo{person}{Peichen Pan}, \bibinfo{person}{Yuxin Wang}, {and} \bibinfo{person}{Peng Zhang}.} \bibinfo{year}{[n.\,d.]}\natexlab{}.
\newblock \showarticletitle{Source-to-{{Source Optimization}} for {{HLS}}}.
\newblock In \bibinfo{booktitle}{\emph{{{FPGAs}} for {{Software Programmers}}}}, \bibfield{editor}{\bibinfo{person}{Dirk Koch}, \bibinfo{person}{Frank Hannig}, {and} \bibinfo{person}{Daniel Ziener}} (Eds.). \bibinfo{publisher}{{Springer International Publishing}}, \bibinfo{pages}{137--163}.
\newblock
\showISBNx{978-3-319-26406-6 978-3-319-26408-0}
\urldef\tempurl%
\url{https://doi.org/10.1007/978-3-319-26408-0_8}
\showDOI{\tempurl}


\bibitem[Conover(1999)]%
        {conover1999practical}
\bibfield{author}{\bibinfo{person}{W.~J. Conover}.} \bibinfo{year}{1999}\natexlab{}.
\newblock \bibinfo{booktitle}{\emph{Practical Nonparametric Statistics} (\bibinfo{edition}{3rd ed} ed.)}.
\newblock \bibinfo{publisher}{Wiley}, \bibinfo{address}{New York}.
\newblock
\showISBNx{978-0-471-16068-7}
\showLCCN{519.5}


\bibitem[Dai et~al\mbox{.}(2018a)]%
        {dai2018fast}
\bibfield{author}{\bibinfo{person}{Steve Dai}, \bibinfo{person}{Yuan Zhou}, \bibinfo{person}{Hang Zhang}, \bibinfo{person}{Ecenur Ustun}, \bibinfo{person}{Evangeline~FY Young}, {and} \bibinfo{person}{Zhiru Zhang}.} \bibinfo{year}{2018}\natexlab{a}.
\newblock \showarticletitle{Fast and accurate estimation of quality of results in high-level synthesis with machine learning}. In \bibinfo{booktitle}{\emph{2018 IEEE 26th Annual International Symposium on Field-Programmable Custom Computing Machines (FCCM)}}. IEEE, \bibinfo{pages}{129--132}.
\newblock


\bibitem[Dai et~al\mbox{.}(2018b)]%
        {quickest}
\bibfield{author}{\bibinfo{person}{Steve Dai}, \bibinfo{person}{Yuan Zhou}, \bibinfo{person}{Hang Zhang}, \bibinfo{person}{Ecenur Ustun}, \bibinfo{person}{Evangeline~F.Y. Young}, {and} \bibinfo{person}{Zhiru Zhang}.} \bibinfo{year}{2018}\natexlab{b}.
\newblock \showarticletitle{Fast and Accurate Estimation of Quality of Results in High-Level Synthesis with Machine Learning}. In \bibinfo{booktitle}{\emph{2018 IEEE 26th Annual International Symposium on Field-Programmable Custom Computing Machines (FCCM)}}. \bibinfo{pages}{129--132}.
\newblock
\urldef\tempurl%
\url{https://doi.org/10.1109/FCCM.2018.00029}
\showDOI{\tempurl}


\bibitem[Ferretti et~al\mbox{.}({[n.\,d.]})]%
        {db4hls}
\bibfield{author}{\bibinfo{person}{Lorenzo Ferretti}, \bibinfo{person}{Jihye Kwon}, \bibinfo{person}{Giovanni Ansaloni}, \bibinfo{person}{Giuseppe Di~Guglielmo}, \bibinfo{person}{Luca Carloni}, {and} \bibinfo{person}{Laura Pozzi}.} \bibinfo{year}{[n.\,d.]}\natexlab{}.
\newblock \bibinfo{booktitle}{\emph{{{DB4HLS}}: {{A Database}} of {{High-Level Synthesis Design Space Explorations}}}}.
\newblock
\urldef\tempurl%
\url{https://doi.org/10.48550/arXiv.2101.00587}
\showDOI{\tempurl}
\showeprint[arxiv]{2101.00587}~[cs]


\bibitem[Gautier et~al\mbox{.}(2016)]%
        {gautier2016spector}
\bibfield{author}{\bibinfo{person}{Quentin Gautier}, \bibinfo{person}{Alric Althoff}, \bibinfo{person}{Pingfan Meng}, {and} \bibinfo{person}{Ryan Kastner}.} \bibinfo{year}{2016}\natexlab{}.
\newblock \showarticletitle{Spector: An opencl fpga benchmark suite}. In \bibinfo{booktitle}{\emph{2016 International Conference on Field-Programmable Technology (FPT)}}. IEEE, \bibinfo{pages}{141--148}.
\newblock


\bibitem[Goswami et~al\mbox{.}({[n.\,d.]})]%
        {mlsbench}
\bibfield{author}{\bibinfo{person}{Pingakshya Goswami}, \bibinfo{person}{Masoud Shahshahani}, {and} \bibinfo{person}{Dinesh Bhatia}.} \bibinfo{year}{[n.\,d.]}\natexlab{}.
\newblock \showarticletitle{{{MLSBench}}: {{A Synthesizable Dataset}} of {{HLS Designs}} to {{Support ML Based Design Flows}}}. In \bibinfo{booktitle}{\emph{Proceedings of the 2020 {{ACM}}/{{SIGDA International Symposium}} on {{Field-Programmable Gate Arrays}}}} ({New York, NY, USA}, 2020-02-24) \emph{(\bibinfo{series}{{{FPGA}} '20})}. \bibinfo{publisher}{{Association for Computing Machinery}}, \bibinfo{pages}{312}.
\newblock
\showISBNx{978-1-4503-7099-8}
\urldef\tempurl%
\url{https://doi.org/10.1145/3373087.3375378}
\showDOI{\tempurl}


\bibitem[Haaswijk et~al\mbox{.}(2018)]%
        {ML_iscas1}
\bibfield{author}{\bibinfo{person}{Winston Haaswijk}, \bibinfo{person}{Edo Collins}, \bibinfo{person}{Benoit Seguin}, \bibinfo{person}{Mathias Soeken}, \bibinfo{person}{Frédéric Kaplan}, \bibinfo{person}{Sabine Süsstrunk}, {and} \bibinfo{person}{Giovanni De~Micheli}.} \bibinfo{year}{2018}\natexlab{}.
\newblock \showarticletitle{Deep Learning for Logic Optimization Algorithms}. In \bibinfo{booktitle}{\emph{2018 IEEE International Symposium on Circuits and Systems (ISCAS)}}.
\newblock


\bibitem[Hara et~al\mbox{.}({[n.\,d.]})]%
        {chstone}
\bibfield{author}{\bibinfo{person}{Yuko Hara}, \bibinfo{person}{Hiroyuki Tomiyama}, \bibinfo{person}{Shinya Honda}, \bibinfo{person}{Hiroaki Takada}, {and} \bibinfo{person}{Katsuya Ishii}.} \bibinfo{year}{[n.\,d.]}\natexlab{}.
\newblock \showarticletitle{{{CHStone}}: A Benchmark Program Suite for Practical {{C-based}} High-Level Synthesis}. In \bibinfo{booktitle}{\emph{2008 {{IEEE International Symposium}} on {{Circuits}} and {{Systems}} ({{ISCAS}})}} (2008-05). \bibinfo{pages}{1192--1195}.
\newblock
\showISSN{2158-1525}
\urldef\tempurl%
\url{https://doi.org/10.1109/ISCAS.2008.4541637}
\showDOI{\tempurl}


\bibitem[Kastner et~al\mbox{.}(2018)]%
        {kastner2018parallel}
\bibfield{author}{\bibinfo{person}{Ryan Kastner}, \bibinfo{person}{Janarbek Matai}, {and} \bibinfo{person}{Stephen Neuendorffer}.} \bibinfo{year}{2018}\natexlab{}.
\newblock \bibinfo{title}{Parallel Programming for {{FPGAs}}}.
\newblock
\newblock
\urldef\tempurl%
\url{https://doi.org/10.48550/arXiv.1805.03648}
\showDOI{\tempurl}
\showeprint[arxiv]{arXiv:1805.03648}


\bibitem[Kim et~al\mbox{.}(2018)]%
        {ml_eda2}
\bibfield{author}{\bibinfo{person}{Ryan~Gary Kim}, \bibinfo{person}{Janardhan~Rao Doppa}, {and} \bibinfo{person}{Partha~Pratim Pande}.} \bibinfo{year}{2018}\natexlab{}.
\newblock \showarticletitle{Machine Learning for Design Space Exploration and Optimization of Manycore Systems}. In \bibinfo{booktitle}{\emph{2018 IEEE/ACM International Conference on Computer-Aided Design (ICCAD)}}. \bibinfo{pages}{1--6}.
\newblock


\bibitem[Lin et~al\mbox{.}(2022)]%
        {PowerGear}
\bibfield{author}{\bibinfo{person}{Zhe Lin}, \bibinfo{person}{Zike Yuan}, \bibinfo{person}{Jieru Zhao}, \bibinfo{person}{Wei Zhang}, \bibinfo{person}{Hui Wang}, {and} \bibinfo{person}{Yonghong Tian}.} \bibinfo{year}{2022}\natexlab{}.
\newblock \showarticletitle{PowerGear: Early-Stage Power Estimation in FPGA HLS via Heterogeneous Edge-Centric GNNs}. In \bibinfo{booktitle}{\emph{Design, Automation \& Test in Europe Conference \& Exhibition (DATE)}}.
\newblock
\urldef\tempurl%
\url{https://doi.org/10.23919/DATE54114.2022.9774682}
\showDOI{\tempurl}


\bibitem[Lin et~al\mbox{.}(2020)]%
        {HL-pow}
\bibfield{author}{\bibinfo{person}{Zhe Lin}, \bibinfo{person}{Jieru Zhao}, \bibinfo{person}{Sharad Sinha}, {and} \bibinfo{person}{Wei Zhang}.} \bibinfo{year}{2020}\natexlab{}.
\newblock \showarticletitle{{HL-Pow: A Learning-Based Power Modeling Framework for High-Level Synthesis}}. In \bibinfo{booktitle}{\emph{25th Asia and South Pacific Design Automation Conference (ASP-DAC)}}.
\newblock
\urldef\tempurl%
\url{https://doi.org/10.1109/ASP-DAC47756.2020.9045442}
\showDOI{\tempurl}


\bibitem[Liu and Schafer(2016)]%
        {ML_DSE}
\bibfield{author}{\bibinfo{person}{Dong Liu} {and} \bibinfo{person}{Benjamin~Carrion Schafer}.} \bibinfo{year}{2016}\natexlab{}.
\newblock \showarticletitle{Efficient and reliable High-Level Synthesis Design Space Explorer for FPGAs}. In \bibinfo{booktitle}{\emph{2016 26th International Conference on Field Programmable Logic and Applications (FPL)}}.
\newblock


\bibitem[Luo et~al\mbox{.}(2023)]%
        {ml_cgra}
\bibfield{author}{\bibinfo{person}{Yixuan Luo}, \bibinfo{person}{Cheng Tan}, \bibinfo{person}{Nicolas~Bohm Agostini}, \bibinfo{person}{Ang Li}, \bibinfo{person}{Antonino Tumeo}, \bibinfo{person}{Nirav Dave}, {and} \bibinfo{person}{Tong Geng}.} \bibinfo{year}{2023}\natexlab{}.
\newblock \showarticletitle{ML-CGRA: An Integrated Compilation Framework to Enable Efficient Machine Learning Acceleration on CGRAs}. In \bibinfo{booktitle}{\emph{2023 60th ACM/IEEE Design Automation Conference (DAC)}}.
\newblock


\bibitem[Mohammadi~Makrani et~al\mbox{.}(2019)]%
        {hls_pyramid}
\bibfield{author}{\bibinfo{person}{Hosein Mohammadi~Makrani}, \bibinfo{person}{Farnoud Farahmand}, \bibinfo{person}{Hossein Sayadi}, \bibinfo{person}{Sara Bondi}, \bibinfo{person}{Sai~Manoj Pudukotai~Dinakarrao}, \bibinfo{person}{Houman Homayoun}, {and} \bibinfo{person}{Setareh Rafatirad}.} \bibinfo{year}{2019}\natexlab{}.
\newblock \showarticletitle{Pyramid: Machine Learning Framework to Estimate the Optimal Timing and Resource Usage of a High-Level Synthesis Design}. In \bibinfo{booktitle}{\emph{2019 29th International Conference on Field Programmable Logic and Applications (FPL)}}.
\newblock


\bibitem[Reagen et~al\mbox{.}({[n.\,d.]})]%
        {machsuite}
\bibfield{author}{\bibinfo{person}{Brandon Reagen}, \bibinfo{person}{Robert Adolf}, \bibinfo{person}{Yakun~Sophia Shao}, \bibinfo{person}{Gu-Yeon Wei}, {and} \bibinfo{person}{David Brooks}.} \bibinfo{year}{[n.\,d.]}\natexlab{}.
\newblock \showarticletitle{{{MachSuite}}: Benchmarks for Accelerator Design and Customized Architectures}. In \bibinfo{booktitle}{\emph{2014 {{IEEE International Symposium}} on {{Workload Characterization}} ({{IISWC}})}} (2014-10). \bibinfo{pages}{110--119}.
\newblock
\urldef\tempurl%
\url{https://doi.org/10.1109/IISWC.2014.6983050}
\showDOI{\tempurl}


\bibitem[Sarkar et~al\mbox{.}(2023a)]%
        {flowgnn}
\bibfield{author}{\bibinfo{person}{Rishov Sarkar}, \bibinfo{person}{Stefan {Abi-Karam}}, \bibinfo{person}{Yuqi He}, \bibinfo{person}{Lakshmi Sathidevi}, {and} \bibinfo{person}{Cong Hao}.} \bibinfo{year}{2023}\natexlab{a}.
\newblock \showarticletitle{{{FlowGNN}}: A Dataflow Architecture for Real-Time Workload-Agnostic Graph Neural Network Inference}. In \bibinfo{booktitle}{\emph{2023 {{IEEE International Symposium}} on {{High-Performance Computer Architecture}} ({{HPCA}})}}. \bibinfo{publisher}{{IEEE}}, \bibinfo{address}{{Montreal, QC, Canada}}, \bibinfo{pages}{1099--1112}.
\newblock
\showISBNx{978-1-66547-652-2}
\urldef\tempurl%
\url{https://doi.org/10.1109/HPCA56546.2023.10071015}
\showDOI{\tempurl}


\bibitem[Sarkar and Hao(2023)]%
        {lightningsim}
\bibfield{author}{\bibinfo{person}{Rishov Sarkar} {and} \bibinfo{person}{Cong Hao}.} \bibinfo{year}{2023}\natexlab{}.
\newblock \showarticletitle{{{LightningSim}}: Fast and Accurate Trace-Based Simulation for High-Level Synthesis}. In \bibinfo{booktitle}{\emph{2023 {{IEEE}} 31st {{Annual International Symposium}} on {{Field-Programmable Custom Computing Machines}} ({{FCCM}})}}. \bibinfo{publisher}{{IEEE}}, \bibinfo{address}{{Marina Del Rey, CA, USA}}, \bibinfo{pages}{1--11}.
\newblock
\showISBNx{979-8-3503-1205-8}
\urldef\tempurl%
\url{https://doi.org/10.1109/FCCM57271.2023.00010}
\showDOI{\tempurl}


\bibitem[Sarkar et~al\mbox{.}(2023b)]%
        {edgemoe}
\bibfield{author}{\bibinfo{person}{Rishov Sarkar}, \bibinfo{person}{Hanxue Liang}, \bibinfo{person}{Zhiwen Fan}, \bibinfo{person}{Zhangyang Wang}, {and} \bibinfo{person}{Cong Hao}.} \bibinfo{year}{2023}\natexlab{b}.
\newblock \showarticletitle{Edge-{{MoE}}: Memory-Efficient Multi-Task Vision Transformer Architecture with Task-Level Sparsity via Mixture-of-Experts}. In \bibinfo{booktitle}{\emph{2023 {{IEEE}}/{{ACM International Conference}} on {{Computer Aided Design}} ({{ICCAD}})}}. \bibinfo{publisher}{IEEE}, \bibinfo{address}{San Francisco, CA, USA}, \bibinfo{pages}{01--09}.
\newblock
\showISBNx{9798350322255}
\urldef\tempurl%
\url{https://doi.org/10.1109/ICCAD57390.2023.10323651}
\showDOI{\tempurl}


\bibitem[Singha et~al\mbox{.}(2022)]%
        {LEAPER}
\bibfield{author}{\bibinfo{person}{Gagandeep Singha}, \bibinfo{person}{Dionysios Diamantopoulosb}, \bibinfo{person}{Juan Gómez-Lunaa}, \bibinfo{person}{Sander Stuijkc}, \bibinfo{person}{Henk Corporaalc}, {and} \bibinfo{person}{Onur Mutlu}.} \bibinfo{year}{2022}\natexlab{}.
\newblock \showarticletitle{{LEAPER: Fast and Accurate FPGA-based System Performance Prediction via Transfer Learning}}. In \bibinfo{booktitle}{\emph{IEEE 40th International Conference on Computer Design (ICCD)}}.
\newblock
\urldef\tempurl%
\url{https://doi.org/10.1109/ICCD56317.2022.00080}
\showDOI{\tempurl}


\bibitem[Sohrabizadeh et~al\mbox{.}({[n.\,d.]})]%
        {autodse}
\bibfield{author}{\bibinfo{person}{Atefeh Sohrabizadeh}, \bibinfo{person}{Cody~Hao Yu}, \bibinfo{person}{Min Gao}, {and} \bibinfo{person}{Jason Cong}.} \bibinfo{year}{[n.\,d.]}\natexlab{}.
\newblock \bibinfo{booktitle}{\emph{{{AutoDSE}}: {{Enabling Software Programmers}} to {{Design Efficient FPGA Accelerators}}}}.
\newblock
\urldef\tempurl%
\url{https://doi.org/10.48550/arXiv.2009.14381}
\showDOI{\tempurl}
\showeprint[arxiv]{2009.14381}~[cs]


\bibitem[Wang et~al\mbox{.}(2021)]%
        {pacmap}
\bibfield{author}{\bibinfo{person}{Yingfan Wang}, \bibinfo{person}{Haiyang Huang}, \bibinfo{person}{Cynthia Rudin}, {and} \bibinfo{person}{Yaron Shaposhnik}.} \bibinfo{year}{2021}\natexlab{}.
\newblock \showarticletitle{Understanding How Dimension Reduction Tools Work: An Empirical Approach to Deciphering t-SNE, UMAP, TriMap, and PaCMAP for Data Visualization}.
\newblock \bibinfo{journal}{\emph{Journal of Machine Learning Research}} \bibinfo{volume}{22}, \bibinfo{number}{201} (\bibinfo{year}{2021}), \bibinfo{pages}{1--73}.
\newblock
\urldef\tempurl%
\url{http://jmlr.org/papers/v22/20-1061.html}
\showURL{%
\tempurl}


\bibitem[Wei et~al\mbox{.}({[n.\,d.]})]%
        {hlsdataset}
\bibfield{author}{\bibinfo{person}{Zhigang Wei}, \bibinfo{person}{Aman Arora}, \bibinfo{person}{Ruihao Li}, {and} \bibinfo{person}{Lizy John}.} \bibinfo{year}{[n.\,d.]}\natexlab{}.
\newblock \showarticletitle{{{HLSDataset}}: Open-Source Dataset for {{ML-assisted FPGA}} Design Using High Level Synthesis}. In \bibinfo{booktitle}{\emph{2023 {{IEEE}} 34th {{International Conference}} on {{Application-specific Systems}}, {{Architectures}} and {{Processors}} ({{ASAP}})}} ({Porto, Portugal}, 2023-07). \bibinfo{publisher}{{IEEE}}, \bibinfo{pages}{197--204}.
\newblock
\showISBNx{9798350346855}
\urldef\tempurl%
\url{https://doi.org/10.1109/ASAP57973.2023.00040}
\showDOI{\tempurl}


\bibitem[Wolf and Glaser(2013)]%
        {yosys}
\bibfield{author}{\bibinfo{person}{Clifford Wolf} {and} \bibinfo{person}{Johann Glaser}.} \bibinfo{year}{2013}\natexlab{}.
\newblock \showarticletitle{Yosys - a Free {{Verilog}} Synthesis Suite}. In \bibinfo{booktitle}{\emph{Proceedings of the 21st {{Austrian Workshop}} on {{Microelectronics}} ({{Austrochip}})}}. \bibinfo{address}{Linz, Austria}.
\newblock


\bibitem[Wu et~al\mbox{.}(2021)]%
        {wu2021ironman}
\bibfield{author}{\bibinfo{person}{Nan Wu}, \bibinfo{person}{Yuan Xie}, {and} \bibinfo{person}{Cong Hao}.} \bibinfo{year}{2021}\natexlab{}.
\newblock \showarticletitle{Ironman: Gnn-assisted design space exploration in high-level synthesis via reinforcement learning}. In \bibinfo{booktitle}{\emph{Proceedings of the 2021 on Great Lakes Symposium on VLSI}}. \bibinfo{pages}{39--44}.
\newblock


\bibitem[Wu et~al\mbox{.}(2022)]%
        {wu2022ironman}
\bibfield{author}{\bibinfo{person}{Nan Wu}, \bibinfo{person}{Yuan Xie}, {and} \bibinfo{person}{Cong Hao}.} \bibinfo{year}{2022}\natexlab{}.
\newblock \showarticletitle{IRONMAN-PRO: Multiobjective design space exploration in HLS via reinforcement learning and graph neural network-based modeling}.
\newblock \bibinfo{journal}{\emph{IEEE Transactions on Computer-Aided Design of Integrated Circuits and Systems}} \bibinfo{volume}{42}, \bibinfo{number}{3} (\bibinfo{year}{2022}), \bibinfo{pages}{900--913}.
\newblock


\bibitem[Zhang et~al\mbox{.}(2020)]%
        {skynet}
\bibfield{author}{\bibinfo{person}{Xiaofan Zhang}, \bibinfo{person}{Haoming Lu}, \bibinfo{person}{Cong Hao}, \bibinfo{person}{Jiachen Li}, \bibinfo{person}{Bowen Cheng}, \bibinfo{person}{Yuhong Li}, \bibinfo{person}{Kyle Rupnow}, \bibinfo{person}{Jinjun Xiong}, \bibinfo{person}{Thomas Huang}, \bibinfo{person}{Honghui Shi}, \bibinfo{person}{Wen-Mei Hwu}, {and} \bibinfo{person}{Deming Chen}.} \bibinfo{year}{2020}\natexlab{}.
\newblock \showarticletitle{{{SkyNet}}: A Hardware-Efficient Method for Object Detection and Tracking on Embedded Systems}.
\newblock \bibinfo{journal}{\emph{Proceedings of Machine Learning and Systems}}  \bibinfo{volume}{2} (\bibinfo{date}{March} \bibinfo{year}{2020}), \bibinfo{pages}{216--229}.
\newblock


\bibitem[Zhou et~al\mbox{.}({[n.\,d.]})]%
        {rosetta}
\bibfield{author}{\bibinfo{person}{Yuan Zhou}, \bibinfo{person}{Udit Gupta}, \bibinfo{person}{Steve Dai}, \bibinfo{person}{Ritchie Zhao}, \bibinfo{person}{Nitish Srivastava}, \bibinfo{person}{Hanchen Jin}, \bibinfo{person}{Joseph Featherston}, \bibinfo{person}{Yi-Hsiang Lai}, \bibinfo{person}{Gai Liu}, \bibinfo{person}{Gustavo~Angarita Velasquez}, \bibinfo{person}{Wenping Wang}, {and} \bibinfo{person}{Zhiru Zhang}.} \bibinfo{year}{[n.\,d.]}\natexlab{}.
\newblock \showarticletitle{Rosetta: A Realistic High-Level Synthesis Benchmark Suite for Software Programmable {{FPGAs}}}. In \bibinfo{booktitle}{\emph{Proceedings of the 2018 {{ACM}}/{{SIGDA International Symposium}} on {{Field-Programmable Gate Arrays}}}} ({New York, NY, USA}, 2018-02-15) \emph{(\bibinfo{series}{{{FPGA}} '18})}. \bibinfo{publisher}{{Association for Computing Machinery}}, \bibinfo{pages}{269--278}.
\newblock
\showISBNx{978-1-4503-5614-5}
\urldef\tempurl%
\url{https://doi.org/10.1145/3174243.3174255}
\showDOI{\tempurl}


\end{thebibliography}

\pagebreak

\onecolumn

\section{Artifact Appendix}
\subsection{Abstract}

The \ourwork~ framework includes multiple software and dataset components, which are available as public open-source releases and artifacts. We briefly outline these components and how to access them. We plan to expand many aspects of our work in the future (e.g., more built-in HLS benchmarks and designs, additional tool flow integrations, enhanced design frontends) and openly encourage contributions and use of \ourwork.

For users strictly interested in running the artifact evaluation to reproduce data and results for various reported case studies, details can be found in \S\ref{sec:artifact-eval-scripts} and at the following repository: \highlightedurl{https://github.com/sharc-lab/hlsfactory-artifact-eval}.

\subsubsection{\ourwork~ Python Library}
\label{sec:python-lib}
The \ourwork~ Python library, \highlightedtt{hlsfactory}, provides APIs and logic for various features including loading HLS designs (locally on disk or from built-in common HLS benchmarks and designs), expanding designs through design space sampling, running parallel tool flows for HLS vendor tools, and extracting+serializing+archiving structured HLS and FPGA tool data (including reports and build artifacts).

\begin{itemize}[noitemsep]
    \item Source code repository: \highlightedurl{https://github.com/sharc-lab/HLSFactory}
    \begin{itemize}
        \item Archived at \highlightedurl{https://zenodo.org/doi/10.5281/zenodo.12989544} (\highlightedtt{DOI: 10.5281/zenodo.12989544})
    \end{itemize}
    \item Documentation: \highlightedurl{https://sharc-lab.github.io/HLSFactory/docs/}
    \begin{itemize}
        \item Archived as part of the source code repository
    \end{itemize}
    \item Install via \highlightedtt{pip}: \highlightedtt{pip install git+https://github.com/sharc-lab/HLSFactory}
    \item Install via \highlightedtt{conda}: \highlightedtt{conda install --channel https://sharc-lab.github.io/HLSFactory/dist-conda hlsfactory}
    \item Install via \highlightedtt{mamba}: \highlightedtt{mamba install --channel https://sharc-lab.github.io/HLSFactory/dist-conda hlsfactory}
\end{itemize}

We highly encourage users to review the documentation for details on the framework, walkthroughs of various demos and case studies with accompanying code and Jupyter Notebooks, and information on how to extend and add new datasets and built-in designs.

\subsubsection{\ourwork's Collection of HLS Benchmarks and Designs}
\label{sec:hls-designs}
One core contribution of this work is the collection of source code for common HLS benchmarks and other open-source and academic HLS designs. We have created design space descriptions and entry point scripts for each design, necessary for the various tool flows supported by our work, and tested our flow on each design.

Our initial release includes designs from the following sources: Polybench \cite{polybench}, MachSuite \cite{machsuite}, Rosetta \cite{rosetta}, CHStone \cite{chstone}, the "Parallel Programming for FPGAs" textbook \cite{kastner2018parallel}, AMD/Xilinx sample HLS designs \cite{xilinx2021basic, xilinx2022vitis}, and various HLS accelerators from Sharc Lab \cite{dgnnbooster,masknet,skynet,edgemoe}. These designs can be found in the \ourwork~ Python library itself under the repository path \highlightedurltext{https://github.com/sharc-lab/HLSFactory/tree/main/hlsfactory/hls_dataset_sources}{hlsfactory/hls\_dataset\_sources}.

A notable feature is that these designs are built into the packaged Python library, available via \highlightedtt{pip} and \highlightedtt{conda}. Users who install \highlightedtt{hlsfactory} can load designs locally without additional downloads. Additionally, users can still load custom designs locally at runtime.

\subsubsection{Pre-Generated Datasets of HLS Synthesized and Implemented Designs}
\label{sec:pregenerated-datasets}
While running our case studies, we ran various end-to-end dataset generations flows. The pre-generated datasets include design sources (with sampled optimization directives if design space sampling is used), HLS synthesized designs (including HLS reports, generated hardware IP, and HLS scheduling and binding data), and, in some runs, FPGA post-implementation reports. These datasets can save users and researchers significant time by providing a dataset fully synthesized and implemented HLS designs with important intermediate artifacts.

We include the following pre-generated datasets: \highlightedtt{Design Space Base Dataset}, \highlightedtt{Design Space Sampled Dataset}, \highlightedtt{Intel Design Flow Dataset}, \highlightedtt{Parallelization Test Dataset}, \highlightedtt{Regression Benchmarking Test Dataset}.

We archive and host these datasets on Zenodo: \highlightedurl{https://zenodo.org/doi/10.5281/zenodo.13117901 } (\highlightedtt{DOI: 10.5281/zenodo.13117901})

\subsubsection{Scripts to Reproduce Case Study Results}
\label{sec:artifact-eval-scripts}
We provide Python scripts to reproduce the results of various case studies, including figures and numerical results. This includes scripts to generate design datasets from scratch and perform the case study analyses. Generating design data requires HLS and FPGA vendor tools and can take over 24 hours for the largest datasets used in this work. Therefore, users can also use the pre-generated datasets from \S\ref{sec:pregenerated-datasets} and only run the required analysis scripts.

The code for running these scripts as an artifact evaluator, along with detailed instructions, is available at the GitHub repository: \highlightedurl{https://github.com/sharc-lab/hlsfactory-artifact-eval}.

This repository is archived at \highlightedurl{https://zenodo.org/doi/10.5281/zenodo.13117886} (\highlightedtt{DOI: 10.5281/zenodo.13117886}).

\subsection{Artifact Check-List (meta-information)}

{\small
\begin{itemize}[noitemsep]
  \item \textbf{Data set}: \highlightedtt{Polybench}, \highlightedtt{MachSuite}, \highlightedtt{Rosetta}, \highlightedtt{CHStone}, \highlightedtt{"Parallel Programming for FPGAs"},\\\highlightedtt{Xilinx/Vitis-HLS-Introductory-Examples}
  \item \textbf{Run-time environment}: \highlightedtt{Linux}, \highlightedtt{Windows}, \highlightedtt{MacOS}
  \item \textbf{Metrics}: \highlightedtt{Runtime}, \highlightedtt{HLS Reported Latency}, \highlightedtt{HLS Reported Resource Usage}, \highlightedtt{Post-Implementation Timing Metrics},\\\highlightedtt{Post-Implementation Resource Usage}, \highlightedtt{Post-Implementation Power Estimation}
  \item \textbf{Output}: \highlightedtt{HLS Synthesis Reports}, \highlightedtt{HLS Synthesized Hardware IP}, \highlightedtt{Post-Implementation Reports}
  \item \textbf{Experiments}: \highlightedtt{"ML Prediction of Post-Implementation Quality-of-Results Metrics"},\\\highlightedtt{"Design Space Coverage"}, \highlightedtt{"Speedup of Fine-Grained Design Parallelism"}, \highlightedtt{"Targeting Different HLS Vendors"},\\\highlightedtt{"Integrating Released Data from Other Works"}, \highlightedtt{"Regression Benchmarking of HLS Synthesis Tools"}
  \item \textbf{Proprietary EDA tools}: \highlightedtt{AMD/Xilinx Vitis HLS 2023.1}, \highlightedtt{AMD/Xilinx Vivado  2023.1},\\\highlightedtt{AMD/Xilinx Vitis HLS 2021.1}, \highlightedtt{AMD/Xilinx Vivado  2121.1},\\\highlightedtt{Intel HLS Compiler (i++) 21.1.0}, \highlightedtt{Intel Quartus Prime 21.1.0}
  \item \textbf{How much disk space required (approximately)?}: $\approx200$ GB
  \item \textbf{How much time is needed to prepare workflow (approximately)?}: $\approx20$ Minutes
  \item \textbf{How much time is needed to complete experiments (approximately)?}: $\approx24$ hours (using 32 cores)
  \item \textbf{Publicly available?}: Yes!
  \item \textbf{Code licenses (if publicly available)?}: \highlightedtt{GNU AGPLv3}
  \item \textbf{Data licenses (if publicly available)?}: \highlightedtt{CC BY-SA 4.0}
  \item \textbf{Archived (provide DOI)?}: Yes! Datasets: [\highlightedtt{10.5281/zenodo.13117901}], Code Repositories: [\highlightedtt{10.5281/zenodo.12989544}, \highlightedtt{10.5281/zenodo.13117886}]
\end{itemize}
}

\subsection{Description}

\subsubsection{How To Access}
The main artifact evaluation code for reproducing results presented in the paper is hosted at this GitHub repository: \highlightedurl{https://github.com/sharc-lab/hlsfactory-artifact-eval}.

\subsubsection{Hardware Dependencies}
No specialized hardware is needed. We recommend a desktop workstation or server with as many cores as possible (for faster parallel dataset generation) and a common Linux-based distrobution (such as Ubuntu).

\subsubsection{Software Dependencies}
\ourwork~ and the artifact evaluation scripts are implemented in Python and require version 3.10 or higher. The \highlightedtt{hlsfactory} library depends on \highlightedtt{pandas}, \highlightedtt{psutil}, \highlightedtt{PyYAML}, \highlightedtt{tqdm}, and \highlightedtt{python-dotenv}. The artifact evaluation scripts additionally require \highlightedtt{pacmap}, \highlightedtt{scikit\_learn}, \highlightedtt{scipy}, \highlightedtt{seaborn}, and \highlightedtt{hlsfactory}.

\subsubsection{Commercial Software Dependencies}
To run Xilinx-based dataset generation flows, AMD/Xilinx’s Vitis HLS and Vivado are required, with most design runs using version \highlightedtt{2023.1}. The regression testing case study requires version \highlightedtt{2021.1}. For Intel-based flows, Intel’s HLS Compiler and Quartus Prime are needed, with version \highlightedtt{21.1.0} required for the Intel design run.

\subsubsection{Datasets}
All the required HLS designs (source code, tool scripts, design space descriptions) are built into the \highlightedtt{hlsfactory} package itself. For more details, refer to \S\ref{sec:hls-designs}.

\subsection{Installation}
Installation of the \highlightedtt{hlsfactory} package (as described in \S\ref{sec:python-lib}) and Python requirements can be done using \highlightedtt{pip} or \highlightedtt{conda} based tools.

\subsection{Experiment Workflow}
For details on running and generating case study results, please refer to the artifact evaluation repository (\S\ref{sec:artifact-eval-scripts}). The process involves obtaining an HLS dataset either by running a dataset generation script or by sourcing a pre-generated dataset from Zenodo. After obtaining the dataset, the user runs a specific case study analysis or visualization script to generate the relevant figures and results. We also specify which case study analyses require which datasets to be run or sourced.

\subsection{Evaluation and Expected Results}
The analysis scripts should produce figures and numerical results similar to those in the paper. The entire workflow is designed to be deterministic, assuming the vendor tools are deterministic. While we have identified most sources of randomness that we allow users to control with a random seed (e.g., random sampling in design space expansion), some elements remain beyond our control, such as \highlightedtt{pacmap}’s fitting, which is not fully deterministic even with \highlightedtt{random\_state} set.

\subsection{Experiment Customization}
Users and evaluators can modify hardcoded parameters in the dataset generation runs or analysis scripts (e.g., random samples for design space expansion, dimensionality reduction parameters). As “proof-of-concept” demos, our case studies allow for modification and extension of \highlightedtt{hlsfactory} to support new data and tools, both locally at runtime and as contributions to the published Python package.

\subsection{Notes}

For more detailed and complete instructions, please refer to the \highlightedtt{README.md} in the artifact evaluation code repository.

\subsection{Methodology}

Submission, reviewing and badging methodology:\\\highlightedurl{https://www.acm.org/publications/policies/artifact-review-and-badging-current},\\\highlightedurl{http://cTuning.org/ae/submission-20201122.html}, \highlightedurl{https://github.com/ml-eda/artifact-evaluation/}.

\let\baselinestretch\OriginalBaselineStretch

\end{document}